\documentclass[11pt]{article}
\usepackage{xcolor}
\usepackage{cite}
\usepackage{multirow}
\usepackage{enumerate}
\usepackage[scale=0.8]{geometry}

\usepackage{graphicx}
\usepackage[subrefformat=parens,labelformat=parens]{subfig}
\usepackage{amsfonts,amsmath,amssymb}
\numberwithin{equation}{section}  % to get equation numbers of the type (1.1), (1.2), (2.1) etc. for sections 1, 2,...

\usepackage{multirow,multicol}
\usepackage{bbm}
\usepackage{xspace}
\usepackage{slashed}
\usepackage{arydshln}

\newcommand{\bZ}{\mathbbm{Z}}

\newcommand{\km}{{T}}

\usepackage{hyperref}

\begin{document}
\begin{center}
{\LARGE {\bf {\Large\bfseries Symmetries of Calabi-Yau Prepotentials with Isomorphic Flops}}}\\[24pt]
{\bf{Andre Lukas$^{a,}$}\footnote{andre.lukas@physics.ox.ac.uk}},
{\bf{Fabian Ruehle$^{b,c,}$}\footnote{f.ruehle@northeastern.edu}}
\bigskip\\[0pt]
\vspace{0.23cm}
${}^a$ {\it 
Rudolf Peierls Centre for Theoretical Physics, University of Oxford\\
Parks Road, Oxford OX1 3PU, UK
}\\[2ex]
${}^b$ {\it 
Department of Physics \& Department of Mathematics, Northeastern University\\
360 Huntington Avenue, Boston, MA 02115, United States
}\\[2ex]
${}^c$ {\it 
The NSF AI Institute for Artificial Intelligence and Fundamental Interactions\\ 
Boston, MA, United States 
}
\end{center}
\vspace{0.5cm}

\begin{abstract}\noindent
Calabi-Yau threefolds with infinitely many flops to isomorphic manifolds have an extended K\"ahler cone made up from an infinite number of individual K\"ahler cones. These cones are related by reflection symmetries across flop walls. We study the implications of this cone structure for mirror symmetry, by considering the instanton part of the prepotential in Calabi-Yau threefolds. We show that such isomorphic flops across facets of the K\"ahler cone boundary give rise to symmetry groups isomorphic to Coxeter groups. In the dual Mori cone, non-flopping curve classes that are identified under these groups have the same Gopakumar-Vafa invariants. This leads to instanton prepotentials invariant under Coxeter groups, which we make manifest by introducing appropriate invariant functions. For some cases, these functions can be expressed in terms of theta functions whose appearance can be linked to an elliptic fibration structure of the Calabi-Yau manifold.
\end{abstract}

\setcounter{footnote}{0}
\setcounter{tocdepth}{2}
\newpage
\tableofcontents

%%%%%%%%%%%%%%%%%%%%%%%%%%%%%%%%%%%%%%%%%%%%%%%%%%%%%%%%%%%%%%%%%%%%%%%%%%%%
\section{Introduction}
Topology change is an interesting and characteristic feature of string theory which has been studied for some time~\cite{Aspinwall:1993yb,Aspinwall:1993nu,Greene:1996cy}, mostly in the context of string compactifications on Calabi-Yau (CY) threefolds. There are two main types of topology changing transitions for CY threefolds: a milder form known as flop transitions and a more severe form, the conifold transitions. In the paper, we will be interested in the former.

Flop transitions of CY threefolds $X$ are known to leave the Hodge numbers $h:=h^{1,1}(X)$ and $h^{2,1}(X)$ unchanged but they can change more refined topological invariants, such as the intersection form and the second Chern class of the tangent bundle. Recently, it has been noted~\cite{Brodie:2021toe,Brodie:2021nit} that topology-preserving flop transition $X\rightarrow X'$ between two isomorphic CY threefolds $X$ and $X'$ are by no means rare. Such isomorphic flops, as we will call them, and their implications are the main topic of this paper.

Suppose an isomorphic flop arises at a boundary facet of the K\"ahler cone $\mathcal{K}$ of $X$. Then, there is an involution, which, relative to a suitable basis $(D_i)$ of divisor classes, can be described by an $h\times h$ matrix $\tilde{\mathcal{M}}$ (satisfying $\tilde{\mathcal{M}}^2=\mathbbm{1}_{h\times h}$). If two divisors $D=k^iD_i$ and $D'={k'}^iD_i$ are related by this involution, that is, $k'=\tilde{\mathcal{M}}k$, then it turns out that their associated linear systems have the same dimension~\cite{Brodie:2020fiq}, so $h^0({\cal O}_X(D))=h^0({\cal O}_X(D'))$. Isomorphic flops can arise across more than one facet of the K\"ahler cone boundary. In this case, we have multiple involutions and corresponding matrices $\tilde{\mathcal{M}}_1,\ldots ,\tilde{\mathcal{M}}_k$ with $\tilde{\mathcal{M}}_1^2=\cdots=\tilde{\mathcal{M}}_k^2=\mathbbm{1}_{h\times h}$, generating groups $\tilde{G}$. Such groups generated by reflections were introduced by Coxeter~\cite{Coxeter} and further studied by Tits and Vinberg~\cite{Vinberg:1971aaa}. Especially the latter studies reflections along the walls of polyhedral cones, and shows that these correspond to Coxeter groups. This structure gives rise to infinite sequences of isomorphic flops generated by repeatedly reflecting the K\"ahler cone along a flop wall and its reflection images under $\tilde{\mathcal{M}}_i$. The union $\mathcal{K}_{\rm ext}=\cup_\alpha\mathcal{K}_\alpha$ of their K\"ahler cones is referred to as the extended K\"ahler cone and is mirror-dual to the complex structure moduli space of the mirror of $X$~\cite{Aspinwall:1993nu}. It turns out, the zeroth cohomology of line bundles on $X$ is invariant under the entire group $\tilde{G}$ and this fact can be immensely helpful for deriving formulae for cohomology~\cite{Brodie:2020fiq}. In the context of infinite flop chains acting on divisors, the authors of~\cite{Gendler:2022qof} recently studied Euclidean D3 branes and noted that theta functions also appear in the non-perturbative superpotential of Type IIB.

In the present paper we are interested in the implications of this symmetry for Gopakumar-Vafa (GV) invariants~\cite{Gopakumar:1998ii,Gopakumar:1998jq} and the instanton prepotential for K\"ahler moduli. To this end, we introduce a (dual) basis $(C^i)$ of curve classes and represent arbitrary classes $C$ by $h$-dimensional integer vectors $d$ such that $C=d_i C^i$. Their GV invariants are denoted by $n_d$. On these curve classes, the involutions act via the matrices $\mathcal{M}_a=\tilde{\mathcal{M}}_a^T$ and the entire group via the dual $G$ of $\tilde{G}$, generated by the matrices $\mathcal{M}_a$. Our main observation is that classes $d$ which do not flop at any of the facets of the (possibly infinite sequence of) CYs $X_\alpha$, the GV invariants are unchanged under the action of $G$, so $n_d=n_{gd}$ for all $g\in G$. This implies that a part of the instanton prepotential for the K\"ahler moduli (specifically, the part associated to non-flopping curve classes) is $\tilde{G}$-invariant and can be expressed in terms of $\tilde{G}$-invariant functions 
\begin{align} 
\label{psidef}
\psi_d^G(T) = \sum_{g\in G} e^{2\pi i (gd)\cdot \km} = \sum_{\tilde{g}\in \tilde{G}}e^{2\pi i d\cdot(\tilde{g}\km)}\qquad\Rightarrow\qquad \psi_d^G(\tilde{g}\km)=\psi_d^G(\km)\quad\forall\; \tilde{g}\in\tilde{G}\,,
\end{align}
where $\km=\chi+it$ are the complexifications of the K\"ahler parameters $t^i$. As far as we are aware, these functions, invariant under certain representations of Coxeter groups, have not been introduced and studied in this context. We will show that, for certain special cases, depending on the underlying CY manifold $X$, they can be expressed in terms of Jacobi theta functions whose appearance can be traced to an elliptic fibration structure of $X$. For explicit examples, we will work with complete intersection CYs in products of projective spaces (CICYs)~\cite{Candelas:1987kf}.

Note added: On the day we submitted our paper to arxiv, a revised version of~\cite{Candelas:2021lkc} appeared on the arxiv, which included a discussion of Coxeter groups in the context of the Hulek-Verrill manifold~\cite{Hulek:2005aaa}.

The plan of the paper is as follows. In Section~\ref{sec:warmup}, we start with a simple warm-up example, a CY manifold with $h^{1,1}(X)=2$ and only a single flop boundary, leading to a finite symmetry $\tilde{G}=\langle \tilde{M}_1\rangle\cong\mathbb{Z}_2$. The insight from this example is used in Section~\ref{sec:Inf-flops_picard2} to study CY manifolds with $h^{1,1}(X)=2$ and two flop boundaries, with the associated groups $\tilde{G}=\langle\tilde{M}_1,\tilde{M}_2\rangle$ isomorphic to universal Coxeter groups with two generators. In Section~\ref{sec:HigherPicardRank} we generalize the discussion to manifolds with $h^{1,1}(X)>2$ but still with two isomorphic flop boundaries and group $\tilde{G}=\langle\tilde{M}_1,\tilde{M}_2\rangle$. As we will show, this case exhibits new features, compared to the $h^{1,1}(X)=2$ case, and, in particular, we find that the $\tilde{G}$ invariant functions $\psi_d^G$ can sometimes be expressed in terms of Jacobi theta functions. The general case with arbitrary $h^{1,1}(X)$ and arbitrary number of flop boundaries, and its relation to Coxeter groups, is discussed in Section~\ref{sec:CoxGroups}. We present our conclusions in Section~\ref{sec:Conclusions}. Appendix~\ref{app:EllFib} contains details about the relation of CYs with infinitely many flops, the occurrence of Jacobi theta functions in the prepotential, and the presence of elliptic fibrations.

%%%%%%%%%%%%%%%%%%%%%%%%%%%%%%%%%%%%%%%%%%%%%%%%%%%%%%%%%%%%%%%%%%%%
\section{A simple warm-up example}
\label{sec:warmup}
We are mostly interested in cases where the group $G$ is of infinite order. However, there are two complications, related to the structure of $G$, in analyzing such cases. First, it is difficult to express elements of Coxeter groups with more than two generators in a sufficiently systematic way in terms of the generators, which we rely on to facilitate computations such as working out the sums in~\eqref{psidef}. Secondly, even for Coxeter groups with only two generators, where writing group elements in terms of generators is relatively simple, the actual matrices in $G$ become complicated. For this reason, we start our discussion with the simple single-flop case where $G\cong\mathbb{Z}_2$, thereby cutting out the above-mentioned complications, and leaving us to focus on other issues, such as the relevant cone structures and properties of GV invariants. Cases with infinite order groups $G$ will be discussed in the subsequent sections.

\subsection{Single-flop cases with Picard rank two}\label{subsec:singleflop}
Consider a CY manifold $X$ with $h:=h^{1,1}(X)=2$, a basis $(D_1,D_2)$ of divisor classes generating the K\"ahler cone, corresponding K\"ahler parameters $t=(t^1,t^2)$ and K\"ahler forms $J_X=t^iD_i$, where $t^i\geq 0$. So, relative to our chosen divisor basis, the K\"ahler cone is then the positive quadrant, $\mathcal{K}=\{t_1e_1+t_2e_2\,|\,t^i\geq 0\}$ (where $e_i$ denote the standard unit vectors). We assume that $\{t^2=0\}$ is a boundary of the effective cone\footnote{The discussion would not change if it was just the boundary of the extended K\"ahler cone.} while $\{t^1=0\}$ is a boundary with an isomorphic flop. Such a structure can be easily detected from the triple intersection numbers $\lambda_{ijk}=D_i\cdot D_j\cdot D_k$ by introducing the quantities
\begin{align}
\label{eq:m1m2def}
 m_1=\frac{2\lambda_{122}}{\lambda_{222}}\,,\qquad m_2=\frac{2\lambda_{112}}{\lambda_{111}}\,.
\end{align}
An isomorphic flop at $\{t^1=0\}$ occurs (for generic complex structure choice) if $m_1$ is finite and integral, and $\{t^2=0\}$ ends the effective cone if $m_2$ does not satisfy these requirements. In this case, passing through $t^1=0$ leads to an isomorphic CY $X'$  and an involution, $\tilde{G}=\langle \tilde{M}_1\rangle\cong\mathbb{Z}_2$ generated by
\begin{align}
\label{eq:SymmetryFlopKahler}
\tilde{M}_1=\begin{pmatrix}
-1&0\\
m_1&1\\
\end{pmatrix}\,.
\end{align}
We reserve the symbol $M$ for the $2\times 2$ reflection matrix and later, when $h^{1,1}(X)>2$, use the symbol $\mathcal{M}$ for higher-dimensional matrices.

The K\"ahler cone $\mathcal{K}'=\mathbb{R}^+(\tilde{v}_1,\tilde{v}_2)$ of $X'$ is generated by
\begin{align}
 \tilde{v}_1=\tilde{M}_1e_1=\left(\begin{array}{r}-1\\m_1\end{array}\right)\,,\qquad
 \tilde{v}_2=\tilde{M}_1e_2=e_2\,,
\end{align}
and the two K\"ahler cones are exchanged by the involution, $\mathcal{K}'=\tilde{M}_1\mathcal{K}$. In particular, since the boundary along $e_1$ is mapped into the one along $v_1$ the latter just as the former must be a boundary of the effective cone, so that $\mathcal{K}_{\rm eff}={\mathcal{K}}\cup{\mathcal{K}}'=\{t_1^\text{eff}\tilde{v}_1+t_2^\text{eff}e_2\,|\,t_i^{\rm eff}\geq 0\}$.

Homology classes $C=d_1C^1+d_2C^2$ are labeled by integer vectors $d=(d_1,d_2)$, relative to a dual basis $(C^i)$, so $C^i\cdot D_j=\delta^i_j$, and their GV invariants are denoted by $n_d$. Curve classes are acted on by the dual group $G=\langle M_1\rangle\cong\mathbb{Z}_2$, with generator
\begin{align}
 M_1=\tilde{M}_1^T=\left(\begin{array}{cc}-1&m_1\\0&1\end{array}\right)\,,
\end{align}
and the dual (Mori) cones\footnote{The Mori cones are denoted by $\mathcal{M}$, $\mathcal{M}'$, etc., while reflections are denoted by $\mathcal{M}_1$, $\mathcal{M}_2$, etc. Despite this slight clash of notation, it should hopefully be clear which object is meant in any given context.}
\begin{align}
 {\mathcal M}=\{c_1 e_1+ c_2 e_2\,|\,c_i\geq 0\}\,,\qquad {\mathcal M}'=\{c'_1v_1+c'_2v_2\,|\,c'_i\geq 0\}\,,\qquad v_1=\left(\begin{array}{c}m_1\\1\end{array}\right)\,,\quad  v_2=-e_1
\end{align}
are exchanged by the action of $M_1$. We call their intersection $\mathcal{M}_{\rm restr}=\mathcal{M}\cap \mathcal{M}'=\{c_1^\text{restr}v_1+c_2^\text{restr}e_2\,|\, c_i^{\rm restr}\geq 0\}$ the restricted cone.

To discuss what happens to GV invariants under the action of $G$, we should distinguish between flopping and non-flopping curve classes.

Non-flopping curve classes are those for which all curves retain a finite volume under a flop transition, that is, all curve classes that are not in the codimension 1 facet of the Mori cone dual to the flop wall of the K\"ahler cone. These must have $G$-invariant GV invariants, $n_{gd}=n_d$ for all $g\in G$, since the number of curves in such a class is not affected by the flop. Non-flopping classes split into two groups, namely those in the restricted cone $\mathcal{M}_{\rm restr}$ and those outside. Under the action of $G$, non-flopping classes in $d\in\mathcal{M}\setminus\mathcal{M}_{\rm restr}$ are mapped outside the Mori cone of $X$ and consequently their GV invariants must vanish. In other words, all non-flopping classes with positive GV invariants reside in the ($G$-invariant) restricted cone. Within the restricted cone, we can identify a fundamental region of the $G$-action, $\mathcal{M}_f\subset\mathcal{M}_{\rm restr}$. For the present case, this can be chosen as $\mathcal{K}_f=\{c_1^fw+c_2^fe_2\,|\,c_i^f\geq 0\}$. We illustrate this cone structure for the example (to be discussed in more detail in the following sub-section) in Figure~\ref{fig:ExampleSingleFlop}.

Flopping curve classes are classes with $n_d>0$ that contain curves shrinking to zero volume, $d_it^i\to0$, at the flop locus. In the present case, the flop wall is $\{t^1=0\}$, so they must be of the form $d=(d_1,0)$. In fact, among those, the only classes which can have non-zero GV invariants~\cite{Brodie:2021toe} are $(1,0)$ and $(2,0)$, so $n_{(d_1,0)}=0$ for $d_1>2$. We collect all flopping classes in a set $\mathcal{B}$, which in this example is simply $\mathcal{B}\subset\{(1,0),(2,0)\}$. The GV invariants for these flopping classes are not $G$-invariant: The image of a curve class $d$ with $n_d\neq 0$ is $M_1(d_1,0)^T=(-d_1,0)^T$, which is outside the cone of effective curves $\mathcal{M}$ and, hence, its GV should vanish. Of course, from the point of view of the manifold $X'$ with K\"ahler parameter $t_1'=-t_1$ the situation is reversed and its GV invariants satisfy $n'_{(d_1,0)}=0$ for all $d_1>0$ and $n'_{(-d_1,0)}>0$ for $d_1\in\{1,2\}$. 

Now we have enough information to discuss the implications of the symmetry $G$ for the instanton prepotential~\cite{Candelas:1990rm}
\begin{align}
\label{eq:InstPrePot}
\mathcal{F}_\text{inst} = \sum_{d}n_d\; \text{Li}_3(e^{2\pi i d\cdot \km})\,.
\end{align}
Splitting the sum into flopping and non-flopping classes and introducing $G$-orbits for the latter we have
\begin{align}
\label{eq:FZ2}
 \mathcal{F}_\text{inst} =\left(\sum_{d\in\mathcal{B}}n_d+\sum_{d\in\mathcal{M}_{\rm restr}}n_d\right)
 \; \text{Li}_3(e^{2\pi i d\cdot \km})=\sum_{d\in\mathcal{B}}n_d\,\text{Li}_3(e^{2\pi i d\cdot \km})+\sum_{d\in \mathcal{M}_f}n_d\Psi_d^G(\km)
\end{align}
where we have introduced the $G$-invariant functions
\begin{align}
\label{eq:PsiZ2}
 \Psi_d^G(\km)=\sum_{g\in G}\text{Li}_3(e^{2\pi i (gd)\cdot \km})=\text{Li}_3(e^{2\pi i d\cdot \km})+\text{Li}_3(e^{2\pi i (M_1d)\cdot \km})\,.
\end{align}
While the first term in Eq.~\eqref{eq:FZ2} for the flopping classes is not invariant\footnote{This part can be made $G$-invariant by extending $\mathcal{F}_\text{inst}$ to the entire effective cone which amounts to replacing the first sum in Eq.~\eqref{eq:FZ2} by $\sum_{d\in\mathcal{B}}n_d\left(\theta(t^1)\text{Li}_3(e^{2\pi i d\cdot \km})+\theta(-t^1)\text{Li}_3(e^{2\pi i (M_1d)\cdot \km}\right)$.} under the action of $\tilde{G}$ on $\km$, the second term for the non-flopping classes is, thanks to the functions $\Psi_d$ satisfying
\begin{align}
 \Psi_d^G(\tilde{g}\km)=\Psi_d^G(\km)\qquad\mbox{for all}\qquad \tilde{g}\in \tilde{G}\,.
\end{align}
Of course, the present situation with a finite group $\tilde{G}\cong\mathbb{Z}_2$ is relatively simple and the resulting constraints on the prepotential rather mild. Things will become significantly more restrictive (and the functions $\Psi_d^G$ will become more interesting) for the cases with symmetries $\tilde{G}$ of infinite order. Before we move on to this we illustrate the single-flop case with an explicit example.

\subsection{Example for a single flop}
\begin{figure}
\centering
\includegraphics[width=.3\textwidth]{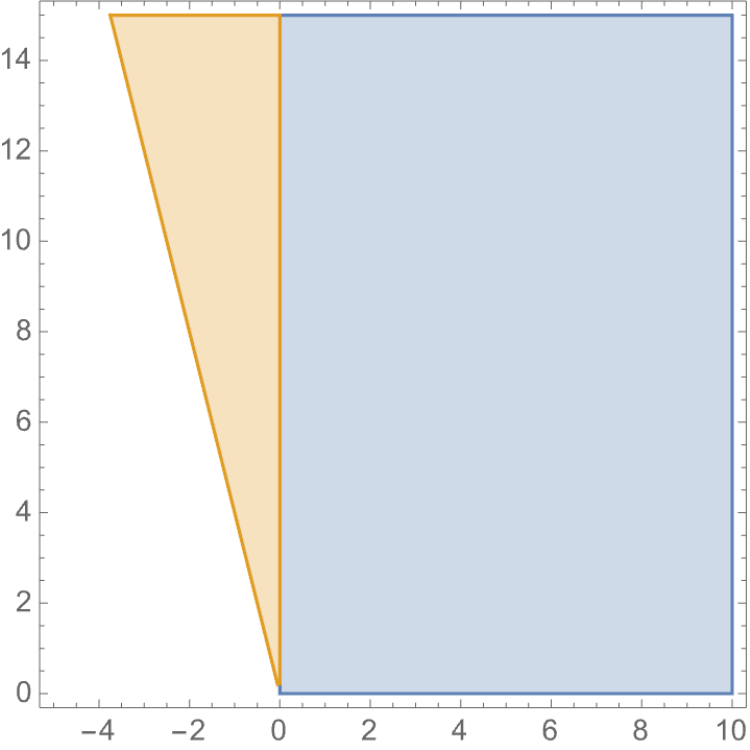}~
\includegraphics[width=.3\textwidth]{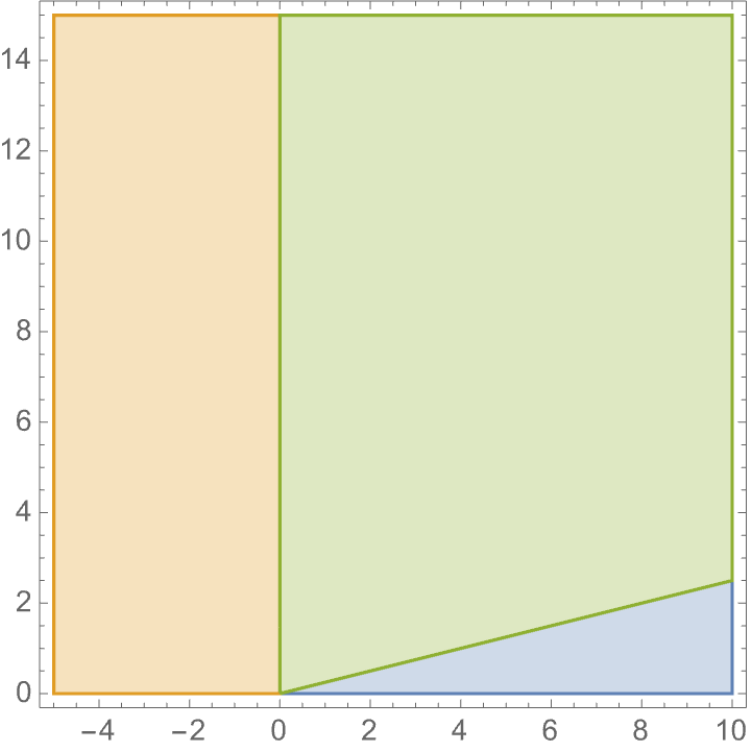}~
\includegraphics[width=.3\textwidth]{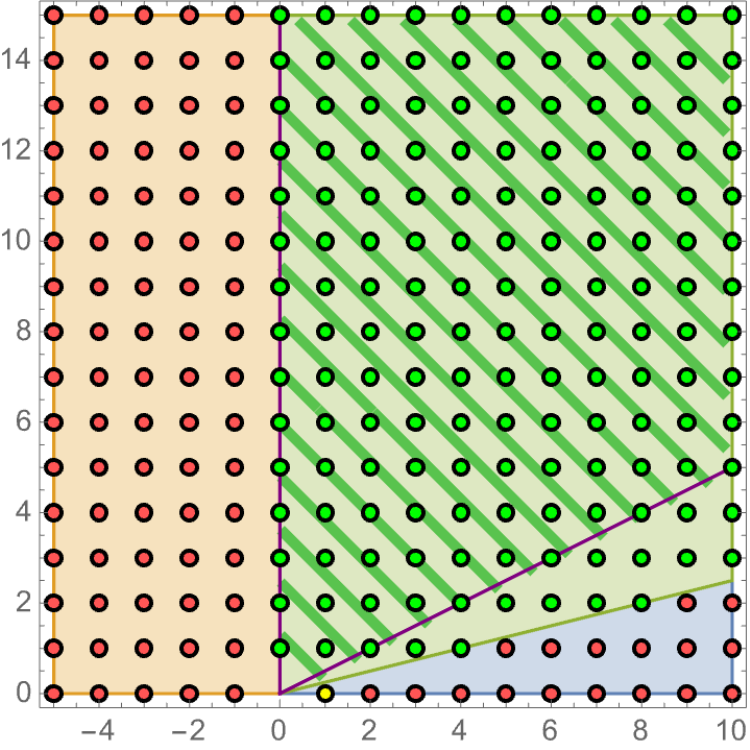}~
\caption{K\"ahler and Mori cone for CICY 7887. \textit{Left}: K\"ahler cones $\mathcal{K}$ (blue) and $\mathcal{K}'$ (orange). The extended cone is $\mathcal{K}_{\text{ext}}=\mathcal{K}\cup\mathcal{K}'$. \textit{Center}: The (dual) Mori cones $\mathcal{M}$ of $\mathcal{K}$ (blue and green) and $\mathcal{M}'$ of $\mathcal{K}'$  (orange and green). Note that the cones overlap in the restricted cone $\mathcal{M}_{\text{restr}}=\mathcal{M}\cap\mathcal{M}'$ (green). \textit{Right}: Similar to the center figure, but we have included the GV invariants as dots. Green dots indicate non-zero GV invariants, red dots indicate zero GV invariants, and yellow dots indicate (non-zero) GV invariants in curve classes that are flopped. A choice for the fundamental cone $\mathcal{M}_f\subset \mathcal{M}_\text{restr}$ is the hatched region between the two purple lines.}
\label{fig:ExampleSingleFlop}
\end{figure}
As an example for the single flop case, we discuss the $h^{1,1}(X)=2$ complete intersection Calabi-Yau manifolds $X$ given as hypersurfaces of degree $(2,4)$ in the ambient space $\mathbbm{P}^1\times\mathbbm{P}^3$ (CICY 7887 in the list of Ref.~\cite{Candelas:1987kf}). Choosing the divisors $(D_1,D_2)$ as the Poincare duals of the standard K\"ahler forms of the ambient projective factors restricted to $X$, the K\"ahler cone relative to this basis is indeed given by the positive quadrant, $\mathcal{K}=\{t_1e_1+t_2e_2\,|\,t^i\geq 0\}$. The non-zero intersection numbers $\lambda_{ijk}=D_i\cdot D_j\cdot D_k$ are $\lambda_{122}=4$ and $\lambda_{222}=2$. Inserting these numbers into Eqs.~\eqref{eq:m1m2def} gives $m_1=4$ while $m_2\not\in\mathbbm{Z}$, which indicates we have indeed a flop boundary at $\{t^1=0\}$ while $\{t^2=0\}$ is a boundary of the effective cone. As a consequence, we have 
\begin{align}
M_1=\begin{pmatrix}
-1&4\\0&1
\end{pmatrix}\,,\qquad
G=\langle\left.M_1~~\right|~~M_1^2=\mathbbm{1}_{2\times2}\rangle\,.
\end{align} 

The resulting K\"ahler cone structure
\begin{align}
 \mathcal{K}=\{t_1e_1+t_2e_2\,|\,t_i\geq 0\}\,,\quad \mathcal{K}'=\{t_1'\tilde{v}_1+t_2' e_2\,|\,t_i'\geq 0\}\,,\quad
 \mathcal{K}_{\rm eff}=\{t^{\rm eff}_1 e_1+t^{\rm eff}_2\tilde{v}_1\,|\,t_i^{\rm eff}\geq 0\}\,,
\end{align}
where $\tilde{v}_1=(-1,4)^T$, is shown in Figure~\ref{fig:ExampleSingleFlop} on the left. For the dual cones this means
\begin{align}
\label{eq:cones7887}
\begin{array}{rclcrcl}
 \mathcal{M}&=&\{c_1 e_1+c_2 e_2\,|\,c_i\geq 0\}&&\mathcal{M}'&=&\{c_1'v_1-c_2'e_1\,|\,c_i'\geq 0\}\\
 \mathcal{M}_{\rm restr}&=&\{c_1^{\rm restr} v_1+ c_2^{\rm restr} e_2\,|\,c_i^{\rm restr}\geq 0\}&& \mathcal{M}_f&=&\{c_1^{f}w+c_1^{f}e_2\,|\,c_i^f\geq 0\}
\end{array} \,,
\end{align}
where $v_1=(4,1)^T$ and $w=(2,1)^T$. The dual cones and the restricted cones are illustrated in Figure~\ref{fig:ExampleSingleFlop} in the middle. The structure of GV invariants is indicated in Figure~\ref{fig:ExampleSingleFlop} on the right. Flopping classes correspond to yellow dots, so for this example there is only a single flopping class, $\mathcal{B}=\{(1,0)\}$, with GV invariant $n_{(1,0)}=64$. Red points indicate classes with vanishing GV invariants while the GV invariants for the green points, contained in the restricted cone, can be positive. The green, hatched region is the fundamental region $\mathcal{K}_f$. Inserting this into Eq.~\eqref{eq:FZ2} with $\Psi_d^G$ defined in Eq.~\eqref{eq:PsiZ2}, we get the instanton prepotential
\begin{align}\label{eq:Fex}
\mathcal{F}_{\rm inst}=64\, {\rm Li}_3\left(e^{2\pi i\km_1}\right)+\sum_{d\in\mathcal{K}_f}n_d \Psi_d^G(\km)=64\, {\rm Li}_3\left(e^{2\pi i\km_1}\right) + \left(\sum_{d\in\mathcal{K}_f}n_d\sum_{g\in\{\mathbbm{1},M_1\}}{\rm Li}_3\left(e^{2\pi i (gd)\cdot T}\right)\right)\,,
\end{align}
where the summation range $\mathcal{K}_f$ is as in Eq.~\eqref{eq:cones7887}.

Out of the $36$ manifolds with $h^{1,1}(X)=2$, defined as complete intersections in product of projective spaces, there are 20, including the above example, with a single flop, as can be seen from the Appendix~A of~\cite{Brodie:2021nit}. There are also $36$ Kreuzer-Skarke CY manifolds with $h^{1,1}(X)=2$ and Appendix~B of~\cite{Brodie:2021nit} shows at least $16$ of those are single-flop cases. All of these manifolds have a prepotential with a structure similar to Eq.~\eqref{eq:Fex}.

%%%%%%%%%%%%%%%%%%%%%%%%%%%%%%%%%%%%%%%%%%%%%%%%%%%%%%%%%%%%%%%%%%%%%%%%%%%%%%%%%%%%%%%%%%%%%%%%%%%%%%%
\section{Infinitely many flops for Picard rank two}
\label{sec:Inf-flops_picard2}
We now turn to manifolds with $h^{1,1}(X)=2$ that admit infinitely many flops. For these manifolds, there exists an infinite symmetry group $G$ and, correspondingly, the orbits of homology classes with identical GV invariants are infinite and the definition of the $G$-invariant functions $\Psi_d^G$ will involve an infinite sum.

The basic set-up and conventions are exactly as in Section~\ref{subsec:singleflop} but, unlike for the single-flop case, we now assume that both $m_1$ and $m_2$ in~\eqref{eq:m1m2def} are integers. In this case, the K\"ahler cone $\mathcal{K}$ has two flop boundaries at $\{t^1=0\}$ and $\{t^2=0\}$ with corresponding involutions generated by
\begin{align}\label{eq:M1M2tilde}
\tilde{M}_1=\left(\begin{array}{cc}-1&0\\m_1&1\end{array}\right)\,,\qquad \tilde{M}_2=\left(\begin{array}{cc}1&m_2\\0&-1\end{array}\right)\,.
\end{align}
These matrices do not commute and lead to a group $\tilde{G}=\langle \tilde{M}_1,\tilde{M}_2\rangle$ as well as its dual $G=\langle M_1,M_2\rangle $ generated by 
\begin{align}\label{eq:M1M2}
M_1=\tilde{M}_1^T=\left(\begin{array}{cc}-1&m_1\\0&1\end{array}\right)\,,\qquad M_2=\tilde{M}_2^T=\left(\begin{array}{cc}1&0\\m_2&-1\end{array}\right)\,.
\end{align}
The structure of $G$ depends on the values of $m_1$ and $m_2$ or, more precisely, on the product $m_1m_2$. There are three cases which lead to qualitatively different results for $G$, namely $m_1m_2<4$, $m_1m_2>4$ and the limiting case $m_1m_2=4$. It turns out only the second case is realized for CY manifolds with $h^{1,1}(X)=2$. In the first case, the effective K\"ahler cone would not be convex, while in the last case, it would be rational polyhedral, that is, it would end with its final wall still containing integral divisors, whose dual curve would, however, not be floppable. Since these inconsistencies do not necessarily occur for $h^{1,1}(X)>2$, it is nevertheless useful to cover all cases.

\subsection{Group structure}
\label{subsec:groupstr2}
To uncover the structure of $G$, it is convenient to introduce the matrices 
\begin{align}\label{eq:STdef}
S=M_1\,,\qquad Q=M_1\,M_2\qquad\Rightarrow\qquad Q^{-1}=SQS=M_2M_1\,.
\end{align}
Then, every group element can be written in the form $Q^kS^\alpha$, where $\alpha\in\{0,1\}$ and $k\in\mathbb{Z}$. Clearly, $S$ has order two and generates a sub-group of $G$ isomorphic to $\mathbb{Z}_2$. On the other hand, the order $N$ of $Q$ can be finite or infinite, depending on the value of $m_1m_2$, and the group $\mathbb{Z}_N$ (with $\mathbb{Z}_\infty=\mathbb{Z}$) it generates is normal in $G$. So in conclusion, we have $G\cong\mathbb{Z}_N\rtimes\mathbb{Z}_2$. It remains to determine the order $N$ and the precise form of the matrices $Q^kS^\alpha$ and to do this we need to distinguish the three cases mentioned above.

\begin{table}[t]
\centering
\begin{tabular}{|c|c|c|}
\hline
$(m_1,m_2)$&order $N$ of $Q$&$G\cong$\\\hline\hline
$(1,1)$&3&$\mathbbm{Z}_3\rtimes\mathbbm{Z}_2$\\
$(1,2)$&4&$\mathbbm{Z}_4\rtimes\mathbbm{Z}_2$\\
$(1,3)$&6&$\mathbbm{Z}_6\rtimes\mathbbm{Z}_2$\\
\hline
\end{tabular}
\caption{Choices for $(m_1,m_2)$ with $m_1m_2<4$ and the resulting group structure.}
\label{tab:finitegroups}
\end{table}

\subsubsection*{The case $\boldsymbol{m_1m_2<4}$}
In these cases the order $N$ of $Q$ is finite, so $G\cong\mathbb{Z}_N\rtimes\mathbb{Z}_2$, and every group element can be written uniquely as $Q^kS^\alpha$, where $k\in\{0,\ldots ,N-1\}$ and $\alpha\in\{0,1\}$. Specifically, we have the three sub-cases listed in Table~\ref{tab:finitegroups}. As mentioned, this case is not realized for CY manifolds with $h^{1,1}(X)=2$, but can occur for $h^{1,1}(X)>2$. We will see an example in Section~\ref{sec:HulekVerrill}.

\subsubsection*{The case $\boldsymbol{m_1m_2> 4}$}
In these cases, the matrix $Q$ has infinite order so that $G\cong\mathbbm{Z}\rtimes\mathbbm{Z}_2$ and every element of $G$ can be uniquely written in the form $Q^kS^\alpha$, where $k\in\mathbbm{Z}$ and $\alpha\in\{0,1\}$. The powers of the matrix $Q$ can be computed explicitly and are given by
\begin{align}
\label{eq:Tk}
Q^k=\left(\begin{array}{cc}a(k)&b(k)\\c(k)&d(k)\end{array}\right)
\end{align}
where 
\begin{align}
\begin{array}{rclcrcl}
a(k)&=&\sum_{i=0}^k (-1)^{k+i}\left(\begin{array}{c}k+i\\2i\end{array}\right)(m_1 m_2)^i\,,&\qquad&d(k)&=&-a(k-1)\,,\\[5mm]
b(k)&=&\sum_{i=1}^k (-1)^{k+i+1}\left(\begin{array}{c}k+i-1\\k-i\end{array}\right) m_1^i m_2^{i-1}\,,&&c(k)&=&-\frac{m_2}{m_1}b(k)\,,
\end{array}
\end{align}
and $d(0)=1$ by definition. Note that the entries of $Q^k$ are power series in $m_i$ whose exponents grow with $k$.

\subsubsection*{The case $\boldsymbol{m_1m_2=4}$}
The boundary case $m_1m_2=4$ is, in a sense, the most interesting one. The order of $Q$ is still infinite, so $G\cong\mathbbm{Z}\rtimes\mathbbm{Z}_2$ and every group element can be uniquely written as $Q^kS^\alpha$, where $k\in\mathbb{Z}$ and $\alpha\in\{0,1\}$.
However, the matrices $Q^k$ behave very differently compared to the case $m_1m_2>4$. Their entries, rather than being power series with exponents growing with $k$, are linear polynomials in $k$. Concretely, for the three possible choices of $(m_1,m_2)$ we find
\begin{align}
\label{eq:Tlimit}
\begin{array}{cclcl}
(m_1,m_2)=(2,2)\;:&& Q^k&=&\left(\begin{array}{cc}2k+1&-2k\\2k&1-2k\end{array}\right)\\[5mm]
(m_1,m_2)=(1,4)\,:&& Q^k&=&\left(\begin{array}{cc}2k+1&-k\\4k&1-2k\end{array}\right)\\[5mm]
(m_1,m_2)=(4,1)\,:&& Q^k&=&\left(\begin{array}{cc}2k+1&-4k\\k&1-2k\end{array}\right)\,.
\end{array}
\end{align}
As mentioned, this case is only realized for $h^{1,1}(X)>2$. As we will see later, the simple form of $Q^k$ for $m_1m_2=4$ and its linear growth with $k$ allows re-writing the $G$-invariant functions $\Psi_d^G$ in terms of Jacobi theta functions.

\subsection{Cones}
For the remainder of this section, we will focus on the case $m_1m_2>4$, the only one actually realized for $h^{1,1}(X)=2$. The group $G$ consists of elements $Q^kS^\alpha$, where $k\in\mathbb{Z}$ and $\alpha\in\{0,1\}$, with the matrices $S$ and $Q$ from Eq.~\eqref{eq:STdef} and it is isomorphic to $\mathbbm{Z}\rtimes\mathbbm{Z}_2$. In addition to the ``central" K\"ahler cone $\mathcal{K}$ of $X$ we have an infinite number of isomorphic CYs with adjacent K\"ahler cones $\mathcal{K}_{\tilde{g}}=\tilde{g}\mathcal{K}$ to either side of $\mathcal{K}$. The union of these forms the effective cone
\begin{align}
 {\mathcal K}_{\rm eff}=\bigcup_{\tilde{g}\in\tilde{G}}\mathcal{K}_{\tilde{g}}=\{t_1^\text{eff}\tilde{v}_1+t_2^\text{eff}\tilde{v}_2\,|\,t_i^{\rm eff}> 0\}\,,\qquad
 \tilde{v}_1=\left(\begin{array}{c}\mu_2\\-1\end{array}\right)\,,\quad
 \tilde{v}_2=\left(\begin{array}{c}-1\\\mu_1\end{array}\right)\,,
\end{align}
where
\begin{align}
 \mu_i=\frac{m_i}{2}\left(1+\sqrt{1-\frac{4}{m_1m_2}}\right)\,.
\end{align}
Note that for the case $m_1m_2>4$ the closure of this cone is irrational. For a specific example, the extended cone is shown in Fig.~\ref{fig:ExampleInfFlop} on the left. The Mori cones dual of the K\"ahler cones $\mathcal{K}_{\tilde{g}}$ are denoted by $\mathcal{M}_g=g\mathcal{M}$ and their intersection, which is the dual of the effective cone, forms the restricted cone
\begin{align}
 \mathcal{M}_{\rm restr}=\bigcap_{g\in G}\mathcal{M}_g=\check{\mathcal K}_{\rm eff}=\{c_1^\text{restr}v_1+c^\text{restr}_2v_2\,|\,c_i\geq 0\}\,,\qquad
 v_1=\left(\begin{array}{c}1\\\mu_2\end{array}\right)\,,\quad v_2=\left(\begin{array}{c}\mu_1\\1\end{array}\right)\,.
\end{align}
An example of a restricted cone is shown in Figure~\ref{fig:ExampleInfFlop}, in the middle.

In our discussion of the GV invariants we will distinguish again between flopping and non-flopping classes. The flopping classes $\mathcal{B}$ of the original manifold $X$ are contained in $\mathcal{B}\subset\{(1,0),(2,0),(0,1),(0,2)\}$, straightforwardly generalizing what we have seen in the single-flop case. As before, for a flopping class $d\in\mathcal{B}$ of $X$, its images $gd$ under the group $G$ are also flopping classes, but rather  for one of the isomorphic manifolds $X'$ instead of $X$. Hence, for every flopping class $d\in\mathcal{B}$ of $X$ we have an entire infinite $G$-orbit of flopping classes (while, for the single flop case where $G\cong\mathbb{Z}_2$ the orbits only had length two). We have already seen that GV invariants are not constant along these orbits. Rather, for a flopping class $d\in\mathbb{B}$ of $X$ we have
\begin{align}
\label{eq:GVflopping}
 n_{gd}=\left\{\begin{array}{cl}n_d&\mbox{for }gd\in\mathcal{M}\\0&\mbox{for }gd\notin\mathcal{M}\end{array}\right. \,.
\end{align}
As before, non-flopping classes $d$ outside the restricted cone always have vanishing GV invariants, $n_d=0$. The restricted cone is $G$-invariant and it splits into $G$-orbits on which the GV invariants are constant, so $n_{gd}=n_d$ for $d\in\mathcal{M}_{\rm restr}$ and all $g\in G$. Within the restricted cone we can choose a fundamental region for the $G$-action as
\begin{align}
 \mathcal{M}_f=\{c_1^fw_1+c_2^fw_2\,|\,c_i^f\geq 0\}\,,\qquad 
 w_1=\left(\begin{array}{c}2\\\mu_2\end{array}\right)\,,\quad v_2=\left(\begin{array}{c}\mu_1\\2\end{array}\right)\,.
\end{align}
For our example, the structure of the various cones and the GV invariants is indicated in Figure~\ref{fig:ExampleInfFlop}, on the right.

\subsection{The prepotential}
Now we have everything in place to write down an expression for the instanton prepotential. As before, we split the sum into flopping and non-flopping classes, using~\eqref{eq:GVflopping} for the former and $G$-invariance of the GV invariants for the latter. This leads to
\begin{align}
\label{eq:FPic2}
 \mathcal{F}_{\rm inst}=\sum_{d\in\mathcal{B}}n_d\sum_{\substack{g\in G\\gd\in\mathcal{M}}}{\rm Li}_3\left(e^{2\pi i (gd)\cdot \km}\right)+\sum_{d\in\mathcal{K}_f}n_d\Psi_d^G(\km)
\end{align}
with the $G$-invariant functions
\begin{align}
\label{eq:PsiPic2}
 \Psi_d^G(\km)=\sum_{g\in G}{\rm Li}_3\left(e^{2\pi i(dg)\cdot \km}\right)=\sum_{k\in\mathbb{Z}}\left({\rm Li}_3\left(e^{2\pi i (Q^kd)\cdot\km}\right)+{\rm Li}_3\left(e^{2\pi i(Q^kSd)\cdot\km}\right)\right)\,.
\end{align}
The functions~\eqref{eq:PsiPic2} contain the matrix powers $Q^k$ from~\eqref{eq:Tk} in their exponents and, hence, these exponents increase with powers of $k$. Functions that behave in this way are not modular -- the exponents for modular functions depend quadratically on $k$. To the best of our knowledge, the functions $\Psi_d^G$ in Eq.~\eqref{eq:PsiPic2} have not been studied before and they do not have an established name. It would be interesting to analyze their properties in more detail.

\subsection{Example with infinitely many flops}
\label{sec:ExamplePicard2}
\begin{figure}[t]
\centering
\includegraphics[width=.3\textwidth]{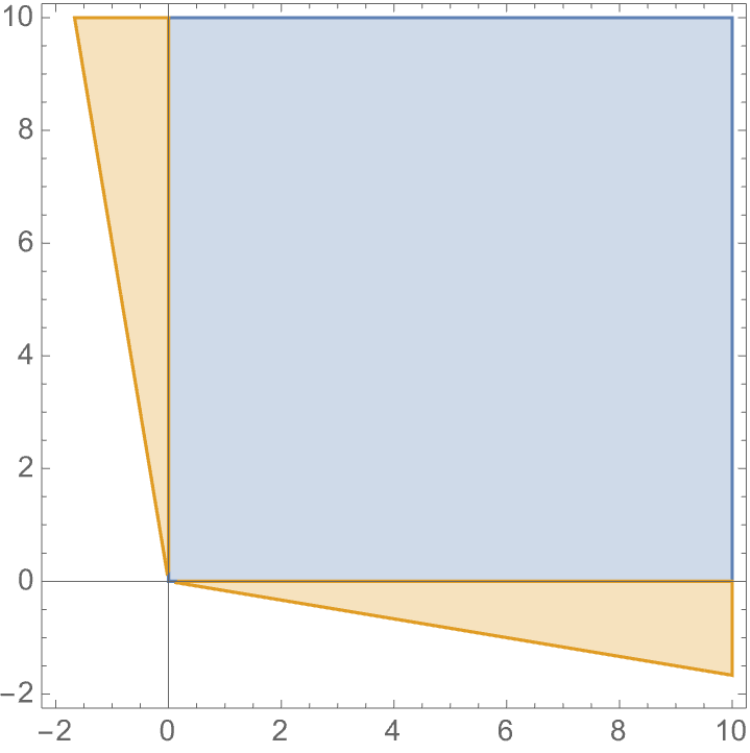}~
\includegraphics[width=.3\textwidth]{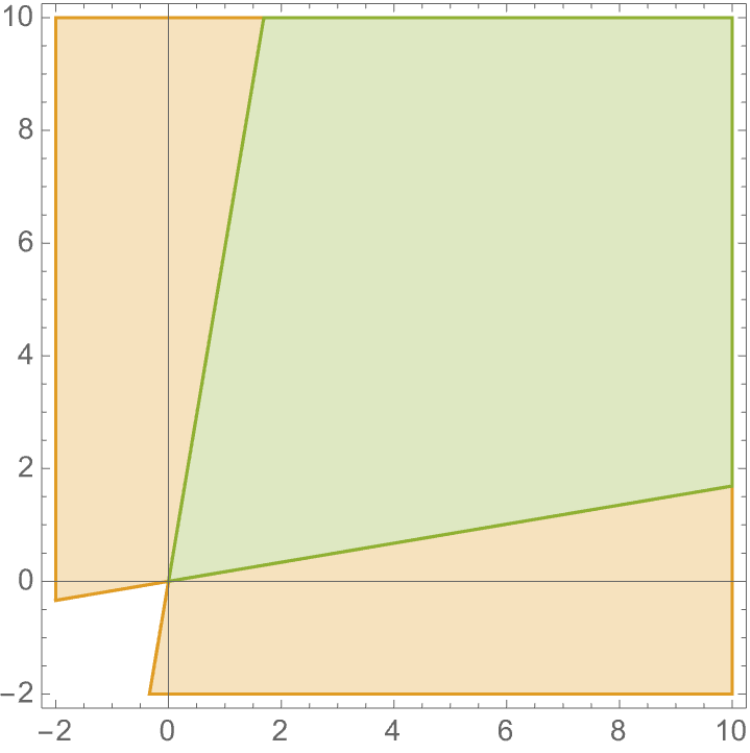}~
\includegraphics[width=.3\textwidth]{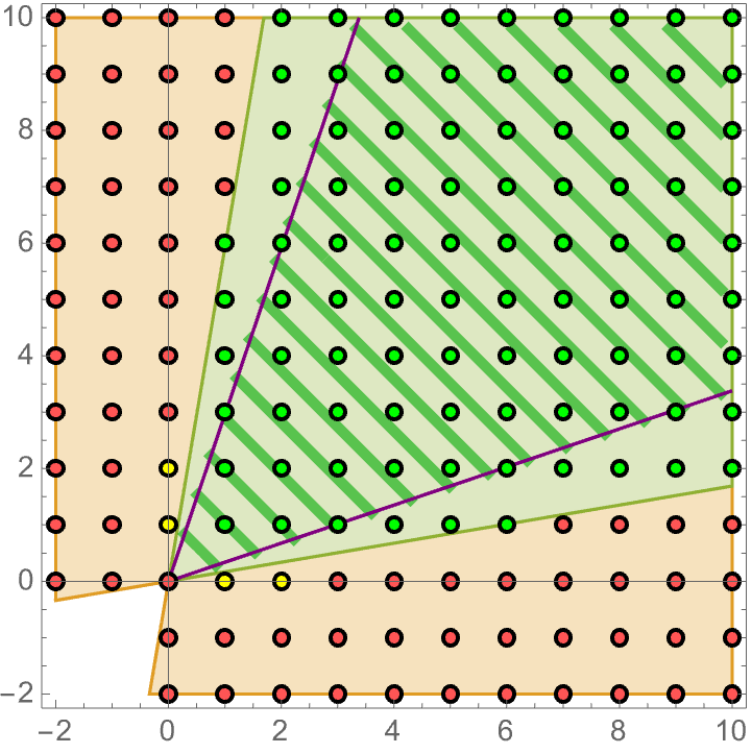}~
\caption{K\"ahler and Mori cone for CICY 7863. \textit{Left}: K\"ahler cone $\mathcal{K}$ (blue) of $X$ and the extended K\"ahler cone $\mathcal{K}_{\text{ext}}$ (blue and orange). \textit{Center}: The (dual) Mori cone $\mathcal{M}$ of $\mathcal{K}$ (blue and green) and the union of all Mori cones $\mathcal{M}'$ of $\mathcal{K}'$  (orange, blue, green).The restricted cone is $\mathcal{M}_{\text{restr}}=\mathcal{M}\cap\mathcal{M}'$ (green). \textit{Right}: We included GV invariants. Green dots indicate non-zero GV invariants, red dots indicate zero GV invariants, and yellow dots indicate (non-zero) GV invariants in curve classes that are flopped. A choice for the fundamental cone $\mathcal{M}_f\subset \mathcal{M}_\text{restr}$ is the hatched region between the two purple lines.}
\label{fig:ExampleInfFlop}
\end{figure}

We would like to illustrate the above structure with a specific example, the complete intersection CY manifold described by the configuration matrix
\begin{align}
X\in \left[\begin{array}{c|ccc}\mathbb{P}^3&2&1&1\\\mathbb{P}^3&2&1&1\end{array}\right]^{2,66}_{-128}\,.
\end{align}
This is CICY 7863 in the standard list of Ref.~\cite{Candelas:1987kf} and it has Hodge numbers $(h^{1,1},h^{2,1})=(2,68)$. The above configuration matrix indicates that the manifold $X$ is defined as the common zero locus of three polynomials in $\mathbb{P}^3\times\mathbb{P}^3$ with bi-degrees $(2,2)$, $(1,1)$ and $(1,1)$. The K\"ahler cone $\mathcal{K}$ (relative to the standard divisor basis $(D_1,D_2)$ associated to pullbacks of the hyperplane classes of the ambient space $\mathbb{P}^3$ factors) is the positive quadrant and the non-zero intersection numbers are given by $\lambda_{111}=\lambda_{222}=2$ and $\lambda_{112}=\lambda_{122}=6$. Equation~\eqref{eq:m1m2def} implies that $m_1=m_2=6$ and hence both boundaries, $\{t^1=0\}$ and $\{t^2=0\}$, exhibit isomorphic flops, resulting in an infinite flop sequence. The various cones are shown in Figure~\ref{fig:ExampleInfFlop}.

The flopping classes of $X$ are given by $\mathcal{B}=\{(1,0),(2,0),(0,1),(0,2)\}$ (indicated by yellow dots in Figure~\ref{fig:ExampleInfFlop} on the right) with $n_{(1,0)}=n_{(0,1)}=80$ and $n_{(2,0)}=n_{(0,2)}=4$. Hence, the first sum in~\eqref{eq:FPic2} runs over four (partial) $G$-orbits. For the elements of $G$ we have
\begin{align}
Q^k=\left(
\begin{array}{cc}
 \frac{\delta _+ \alpha _+^k-\delta _- \alpha _-^k}{4 \sqrt{2}} & \frac{\alpha _-^k-\alpha _+^k}{4
   \sqrt{2}} \\
 \frac{\alpha _+^k-\alpha _-^k}{4 \sqrt{2}} & \frac{\delta _+ \alpha _-^k-\delta _- \alpha _+^k}{4
   \sqrt{2}} \\
\end{array}
\right)\,,\quad
Q^kS=\left(
\begin{array}{cc}
 \frac{\delta _- \alpha _-^k-\delta _+ \alpha _+^k}{4 \sqrt{2}} & \frac{\alpha _+^{k+1}-\alpha
   _-^{k+1}}{4 \sqrt{2}} \\
 \frac{\alpha _-^k-\alpha _+^k}{4 \sqrt{2}} & \frac{\delta _+ \alpha _+^k-\delta _- \alpha _-^k}{4
   \sqrt{2}} \\
\end{array}
\right)\,,
\end{align}
where we have defined
\begin{align}
 \alpha_\pm=17\pm 12\sqrt{2}\,,\qquad \beta_\pm=4\pm 3\sqrt{2}\,,\qquad \gamma_\pm=24\pm 17\sqrt{2}\, ,\quad
 \delta_\pm=3\pm 2\sqrt{2}\,.
\end{align}
Inserting this into~\eqref{eq:PsiPic2}, we get the the $G$-invariant functions $\Psi_d^G$
\begin{align}
\label{eq:Psiexample2}
 \Psi_d^G(\km)=&\sum_{k\in\mathbb{Z}}{\rm Li}_3\left(e^{\frac{1}{4} i \pi  \left(\alpha _-^k \left(d_2 \left(\beta _+ \km _2+\sqrt{2} \km
   _1\right)-d_1 \left(\sqrt{2} \km _2-\beta _- \km _1\right)\right)+\alpha _+^k \left(d_1
   \left(\beta _+ \km _1+\sqrt{2} \km _2\right)-d_2 \left(\sqrt{2} \km _1-\beta _- \km
   _2\right)\right)\right)}\right)+\nonumber\\
     &\sum_{k\in \mathbb{Z}}{\rm Li}_3\left(e^{\frac{1}{4} i \pi  \left(\alpha _-^k \left(d_2 \left(\beta _- \km _2+\gamma _- \km
   _1\right)+d_1 \left(\sqrt{2} \km _2-\beta _- \km _1\right)\right)+\alpha _+^k \left(d_2
   \left(\beta _+ \km _2+\gamma _+ \km _1\right)-d_1 \left(\beta _+ \km _1+\sqrt{2} \km
   _2\right)\right)\right)}\right)
\end{align}
As can be seen from Appendix A of Ref.~\cite{Brodie:2021nit}, among the $h^{1,1}(X)=2$ CICYs, there are six manifolds with infinite flop sequences (including the above example) which realize the values
\begin{align}
 (m_1,m_2)\in\{(4,4),(3,8),(5,8),(6,6),(7,7)\}\,.
\end{align}
They can be analyzed along the same lines as the example above.

%%%%%%%%%%%%%%%%%%%%%%%%%%%%%%%%%%%%%%%%%%%%%%%%%%%%%%%%%%%%%%%%%%%%%%%%%%%%
\section{Manifolds with \texorpdfstring{$\boldsymbol{h^{1,1}(X)>2}$}{h11(X)>2} and two symmetry generators}
\label{sec:HigherPicardRank}
We would now like to generalize our discussion to manifolds with Picard number $h=h^{1,1}(X)>2$. As usual, we introduce a basis $(D_1,\ldots ,D_h)$ of divisor classes generating the K\"ahler cone (assuming, for simplicity, that the K\"ahler cone is simplicial) and we write K\"ahler forms as $J_X=t^iD_i$, with K\"ahler parameters $t^i\geq0$. The presence of isomorphic flop boundaries can then be detected from the intersection numbers $\lambda_{ijk}=D_i\cdot D_j\cdot D_k$ as follows. An isomorphic flop across the K\"ahler cone boundary $\{t^i=0\}$ occurs iff there is an integer vector $u_i$ with components $u_i^a$, $a=1,2,\ldots,h$, that satisfies
\begin{align}
\label{eq:ReflectionVector}
 u_i^i=2\,,\qquad 
 u_i^a\leq0\text{~~~for~~~} a\neq i\,,\qquad
 d_{abc}u_i^a=0\text{~~~for all~~~}\, b,c\neq i\,.
\end{align} 
If such a vector exists then we have an involution acting ``across" the boundary $\{t^i=0\}$ which is generated by
\begin{align}
 \label{eq:tildeMReflect}
 \tilde{\mathcal{M}}_i=\mathbbm{1}_h-(\vec{0},\ldots ,\vec{0},u_i,\vec{0},\ldots ,\vec{0})\,,\qquad\Rightarrow\qquad \mathcal{M}_i=\mathcal{\tilde{M}}^T_i=\mathbbm{1}_h-\begin{pmatrix}\vec{0}^{\,T}\\\vdots\\\vec{0}^{\,T}\\u_i\\\vec{0}^{\,T}\\\vdots\end{pmatrix}
\end{align}
where the vector $u_i$ in the second $h\times h$ matrix appears in the $i^{\rm th}$ row. The symmetry group $\tilde{G}$ is then generated by all such matrices, $\tilde{G}=\langle\tilde{\mathcal{M}}_i\rangle$, while the matrices $\mathcal{M}_i$ generate the dual $G=\langle \mathcal{M}_i\rangle$. It can be shown that matrices of the type~\eqref{eq:tildeMReflect} are indeed reflections, that is, they satisfy $\tilde{\mathcal{M}}_i^2=\mathbbm{1}_{h\times h}$ and they have precisely one eigenvector with eigenvalue $-1$ and $(h-1)$ eigenvectors with eigenvalues $+1$, such that $\det{\tilde{\mathcal{M}}_i}=-1$. The integers $m_1$ and $m_2$ introduced previously are given by $m_1=-u_1^2$ and $m_2=-u_2^1$.

For a manifold with Picard number $h$ (and a simplicial K\"ahler cone) we can have at most $h$ isomorphic flop boundaries and corresponding involutions. They are all isomorphic to Coxeter groups as we will discuss in section~\ref{sec:CoxGroups}. In this section, we will focus on the case of just two involutions. This can arise either because the manifold in question only has two isomorphic flop boundaries or, if there are three or more, because we focus on the Coxeter subgroup generated by just two involutions. It might seem that this situation is not significantly different from the two generator case for $h^{1,1}(X)=2$ we have discussed in the previous section. However, it turns out interesting new features do arise since all of the cases for $m_1m_2$ listed in Section~\ref{subsec:groupstr2} can actually be realized for appropriate choices of manifolds. In particular, there are examples for the interesting limiting case $m_1m_2=4$ where the elements of $G$ have a particularly simple form, cf.~Eq.~\eqref{eq:Tlimit}. As we will see, for such cases the $G$-invariant functions $\Psi_d^G$ can be expressed in terms of Jacobi theta functions.

\subsection{Group structure}
\label{sec:hGenericGroupStructure}
We can assume that the two flop boundaries are $\{t^1=0\}$ and $\{t^2=0\}$, so that the corresponding generators can be written as
\begin{align}
\label{eq:M1M2gen}
\begin{array}{rclcrcl}
 \mathcal{M}_1&=&\left(\begin{array}{c:c}M_1&U_1\\\hdashline0&\mathbbm{1}_{(h-2)\times(h-2)}\end{array}\right)&&
 \mathcal{M}_2&=&\left(\begin{array}{c:c}M_2&U_2\\\hdashline0&\mathbbm{1}_{(h-2)\times(h-2)}\end{array}\right)\\[5mm]
 U_1&=&\left(\begin{array}{l}u_1\\\vec{0}^{\,T}\end{array}\right)&&U_2&=&\left(\begin{array}{l}\vec{0}^{\,T}\\u_2\end{array}\right)
\end{array} \,.
\end{align}
These $(h\times h)$ matrices $\mathcal{M}_i$ are block matrices consisting of a $2\times2$ block $M_i$ from  Eq.~\eqref{eq:M1M2}, a $2\times(h-2)$ block $U_i$, an $(h-2)\times2$ block with the zero-matrix, and a $(h-2)\times(h-2)$ identity matrix. The $1\times (h-2)$-dimensional row vectors $u_1$ and $u_2$ are defined in terms of the GLSM charges as explained in Ref.~\cite{Brodie:2021toe} and reviewed in Appendix~\ref{app:EllFib}. If we define new generators
\begin{align}
 \mathcal{S}=\mathcal{M}_1\,,\qquad \mathcal{Q}=\mathcal{M}_1\mathcal{M}_2=\left(\begin{array}{c:c}Q&U\\\hdashline0&\mathbbm{1}_{(h-2)\times(h-2)}\end{array}\right)\,,\qquad
 U=M_1U_2+U_1=\left(\begin{array}{c}u_1+m_1u_2\\u_2\end{array}\right)
\end{align}
the elements of $G=\langle\mathcal{M}_1,\;\mathcal{M}_2\rangle$ can be uniquely written as $\mathcal{Q}^k\mathcal{S}^\alpha$, where $k\in\mathbbm{Z}$ and $\alpha\in\{0,1\}$. More explicitly, these matrices are
\begin{align}
\label{eq:group2}
 \mathcal{Q}^k=\left(\begin{array}{c:c}Q^k&R_kU\\\hdashline0&\mathbbm{1}_{(h-2)\times(h-2)}\end{array}\right)\,,\;\;
 \mathcal{Q}^k\mathcal{S}=\left(\begin{array}{c:c}Q^kS&Q^kU_1+R_kU\\\hdashline0&\mathbbm{1}_{(h-2)\times(h-2)}\end{array}\right)\,,
\end{align}
where
\begin{align}
 R_k=\left\{\begin{array}{rl}\sum_{j=0}^{k-1}Q^j&\mbox{for }k\geq 0\\-\sum_{j=k}^{-1}Q^j&\mbox{for }k<0\end{array}\right.\,,\qquad
 Q=M_1M_2\,,\qquad S=M_1\,,
\end{align}
and the powers $Q^k$ of $Q$ have already been computed in Section~\ref{subsec:groupstr2}. For the $k<0$ case, note that $M_i u_i=-u_i$ and $M_i^2=\mathbbm{1}_{2\times2}$. The interesting observation is that the qualitative behavior of these matrices with $k$ is controlled by the behavior of the $2\times 2$ matrices $M_1$, $M_2$ and, hence, by the product $m_1m_2$. This means we can essentially use the classification from Section~\ref{subsec:groupstr2}.

\subsubsection*{The case $\boldsymbol{m_1m_2=0}$}
One can show that $m_1=1$ iff $m_2=0$. Then $Q$ as well as $\mathcal{Q}$ are of order two, so that $G\cong\mathbbm{Z}_2\rtimes\mathbbm{Z}_2$. This can only happen for direct product manifolds. 

\subsubsection*{The case $\boldsymbol{0<m_1m_2<4}$}
We have the choices $(m_1,m_2)\in\{(1,1),(1,2),(1,3),(2,1),(3,1)\}$ and it can be checked by explicit computation of the matrices $R_k$, that the order of $\mathcal{Q}$ remains the same as the order of $Q$ for all cases. This means we obtain the same groups as in Table~\ref{tab:finitegroups}. As we explain in Appendix~\ref{app:EllFib}, the cases $(1,2)$ and $(1,3)$ are not realized, at least not by complete intersections in projective spaces.

\subsubsection*{The case $\boldsymbol{m_1m_2>4}$}
We know that the order of $Q$ is infinite for this case, which implies the same is true for $\mathcal{Q}$. This means that the group structure is $G\cong\mathbb{Z}\rtimes\mathbb{Z}_2$. Further, it is clear that (at least some of) the entries of $\mathcal{Q}$ increase as powers of $k$ so that the associated $G$-invariant functions $\Psi_d^G$ are complicated, non-modular functions analogous to the example in~\eqref{eq:Psiexample2}.

\subsubsection*{The case $\boldsymbol{m_1m_2=4}$}
In this case, the order of $Q$ is infinite so that $G\cong\mathbb{Z}\rtimes\mathbb{Z}_2$, but we know from~\eqref{eq:Tlimit} that the matrix elements of $Q^k$ only grow linearly with $k$. Computing the matrices $R_k$,
\begin{align}
\label{eq:Rlimit}
\begin{array}{clcrcrcl}
(m_1,m_2)=(2,2)\;:& Q^k&=&\left(\begin{array}{cc}2k+1&-2k\\2k&1-2k\end{array}\right)\,,&&
 R_k&=&\left(
\begin{array}{cc}
 k^2 & k(1-k) \\
 k (k-1) & k(2-k) \\
\end{array}
\right)\,,\\[5mm]
(m_1,m_2)=(1,4)\,:& Q^k&=&\left(\begin{array}{cc}2k+1&-k\\4k&1-2k\end{array}\right)\,,&&R_k&=&\left(
\begin{array}{cc}
 k^2 & \frac{k}{2} (1-k) \\
 2k (k-1) & k(2-k) \\
\end{array}
\right)\,,\\[5mm]
(m_1,m_2)=(4,1)\,:& Q^k&=&\left(\begin{array}{cc}2k+1&-4k\\k&1-2k\end{array}\right)\,,&&R_k&=&\left(
\begin{array}{cc}
 k^2 & 2k (1-k) \\
 \frac{k}{2} (k-1) & k(2-k) \\
\end{array}
\right)\,,
\end{array}
\end{align}
we see that the entries of $\mathcal{Q}^k$ are at most quadratic polynomials in $k$. This feature means that the functions $\Psi_d^G$ can be expressed in terms of theta functions, as we will show below.

\subsection{The prepotential}
The general structure of the prepotential is given by the same equation~\eqref{eq:FPic2} as for Picard number two, with the obvious generalizations of the various quantities involved. The cases $0\leq m_1m_2<4$ lead to a finite symmetry group $G$ (at least when focusing on a rank 2 subgroup), so that the functions $\Psi_d^G$ are given by a finite sum of exponential terms similar to~\eqref{eq:FZ2}. Likewise, we will not discuss the case $m_1m_2>4$ explicitly -- we have already encountered this case for $h^{1,1}(X)=2$ and it it leads to complicated functions $\Psi_d^G$ similar to~\eqref{eq:Psiexample2}. Instead, we focus on the structure of the $G$-invariant functions $\Psi_d^G$ for $m_im_j=4$. In this case, the group $G\cong\mathbb{Z}_2\rtimes\mathbb{Z}$ is infinite and the matrices in $G$ depend (at most) quadratically on $k$, as Eq.~\eqref{eq:Rlimit} shows.

In general, expanding the tri-logarithm and using Eq.~\eqref{psidef}, the $G$-invariant functions can be written as
\begin{align}
\label{eq:psidef1}
 \Psi(\km)=\sum_{l=1}^\infty\frac{1}{l^3}\psi_{ld}^G(\km)\,,\qquad
 \psi_{ld}^G(\km)=\sum_{g\in G} e^{2\pi il(gd)\cdot\km}=\sum_{k\in\mathbbm{Z}}\left(e^{2\pi i l (\mathcal{Q}^kd)\cdot\km}+e^{2\pi il(\mathcal{Q}^k\mathcal{S}d)\cdot\km}\right)\,.
\end{align}
Of course, the functions $\psi_d^G$ are also $\tilde{G}$-invariant, that is, they satisfy $\psi_d^G(\tilde{g}\km)=\psi_d^G(\km)$ for all $\tilde{g}\in\tilde{G}$. To re-write these expressions, we split all vectors into directions parallel and perpendicular to the two flop directions $i=1,2$, so we write
\begin{align}
 \km=\left(\begin{array}{l}\km_\parallel\\\km_\perp\end{array}\right)\text{~~~with~~~}
 \km_\parallel=\left(\begin{array}{l}\km_1\\\km_2\end{array}\right)\,,\quad\quad
 d=\left(\begin{array}{l}d_\parallel\\d_\perp\end{array}\right)\text{~~~with~~~}
 d_\parallel=\left(\begin{array}{l}d_1\\d_2\end{array}\right)\,.
\end{align}
From~\eqref{eq:group2}, the exponents in~\eqref{eq:psidef1} can then be written as
\begin{align}
\begin{split}
 (\mathcal{Q}^kd)\cdot\km&=(Q^k d_\parallel+R_k U d_\perp)\cdot\km_\parallel+d_\perp\cdot\km_\perp\\
 (\mathcal{Q}^k\mathcal{S}d)\cdot\km&=(Q^k S d_\parallel+(R_k U+Q^k U_1) d_\perp)\cdot\km_\parallel+d_\perp\cdot\km_\perp\,
\end{split}
\end{align}
Defining $\delta_i:=u_i\cdot d_\perp$ and using the matrices from~\eqref{eq:Rlimit}, we find for the case $(m_1,m_2)=(2,2)$
\begin{align}
\label{eq:exp2}
(\mathcal{Q}^k d)\cdot\km=\frac{1}{2}k^2\tau+k z+d\cdot\km\,,\qquad
 (\mathcal{Q}^k \mathcal{S} d)\cdot\km=\frac{1}{2}k^2\tau+k\tilde{z}+y+d\cdot \km\,,
\end{align} 
where
\begin{align}
\label{eq:tauzdef22}
\begin{array}{rl}
 \tau&=2(\delta_1+\delta_2)(\km_1+\km_2)\\
 z&=2(d_1-d_2)(\km_1+\km_2)+\delta_2\km_1-\delta_1\km_2\\
 \tilde{z}&=-z+2(\delta_1+\delta_2) \km_1\\
 y&=(\delta_1-2 d_1+2d_2)\km_1\,.
 \end{array}
\end{align}	
Using these results in Eq.~\eqref{eq:psidef1}, we find that the $G$-invariant functions $\psi_d^G$ can be re-written as
\begin{align}
\label{eq:theta2}
 \psi_d^G(\km)=e^{2\pi id\cdot\km}\vartheta(z;\tau)+e^{2\pi i(d\cdot\km+y)}\vartheta(\tilde{z};\tau)\,.
\end{align}
where
\begin{align}
\vartheta(z;\tau):=\sum_{k\in \mathbb{Z}}e^{\pi i k^2\tau+2\pi i k z}\,.
\end{align}
is the Jacobi theta function.

A similar calculation for $(m_1,m_2)=(1,4)$, using the appropriate matrices from Eq.~\eqref{eq:Tlimit}, gives a result with the same structure as in~\eqref{eq:exp2} and~\eqref{eq:theta2}, but with 
\begin{align}
\label{eq:tauzdef14}
\begin{array}{rl}
 \tau&=(2\delta_1+\delta_2)(\km_1+2\km_2)\,,\\
 z&=(2d_1-d_2)(\km_1+2\km_2)+\frac{1}{2}\delta_2\km_1-2\delta_1\km_2\,,\\
 \tilde{z}&=-z+(2\delta_1+\delta_2)\km_1\,,\\
 y&=(-2d_1+d_2+\delta_1)\km_1\,.
 \end{array}
\end{align}
Of course, the case $(m_1,m_2)=(4,1)$ leads to a similar result as the present case $(m_1,m_2)=(1,4)$, but with the exchange $1\leftrightarrow 2$ of indices on all quantities. Hence, for groups $G$ with two generators and $m_1m_2=4$, we have shown that the $G$-invariant functions which make up the prepotential can be expressed in terms of Jacobi theta functions. 

In order to get a better understanding for why the invariants $\psi^G_d$ can be written in terms of the Jacobi theta function it is useful to translate the action of $\tilde{G}=\langle\tilde{\mathcal{S}},\tilde{\mathcal{Q}}\rangle$ on $T$ to our new variables $\tau$, $z$, $\tilde{z}$, $y$ and $d\cdot T$ defined above, in Eqs.~\eqref{eq:tauzdef22} and~\eqref{eq:tauzdef14}. In particular, we would like to know how the $\tilde{G}$ action relates to the symmetry properties of the theta function, that is, to $\text{SL}(2,\mathbbm{Z})$ transformations and quasi-periodicity. Writing $\tilde{\mathcal S}=\tilde{\mathcal M}_1$ and $\tilde{\mathcal Q}=\tilde{\mathcal M}_1\tilde{\mathcal M}_2$, where $\tilde{\mathcal M}_i=\mathcal{M}_i^T$ with the matrices~\eqref{eq:M1M2gen}, leads for all cases where $m_1m_2=4$ to the $\tilde{\mathcal S}$ and $\tilde{\mathcal Q}$ actions
\begin{align}
\label{eq:ModTrafosFromSQ}
 \tilde{\mathcal S}:
 \left\{\begin{array}{rcl}
 d\cdot T&\mapsto& d\cdot T+y\\
 \tau&\mapsto& \tau\\
 z&\mapsto&-\tilde{z}\\
 \tilde{z}&\mapsto&-z\\
 y&\mapsto&-y
 \end{array}\right.\,,
 \qquad
 \tilde{\mathcal Q}:
 \left\{\begin{array}{rcl}
 d\cdot T&\mapsto& d\cdot T+\frac{1}{2}\tau-z\\
 \tau&\mapsto&\tau\\
 z&\mapsto&z-\tau\\
 \tilde{z}&\mapsto&\tilde{z}-\tau\\
 y&\mapsto&y+z-\tilde{z}\,.
 \end{array}\right.
\end{align}
This shows that the $\tilde{\mathcal S}$ invariance of $\psi_d^G$ arises by a swap of the two theta function terms in Eq.~\eqref{eq:theta2} accompanied by the $\text{SL}(2,\mathbbm{Z})$ transformation $-\mathbbm{1}_2$. Hence, the $\mathbb{Z}_2$ subgroup of $\tilde{G}$ generated by $\tilde{\mathcal S}$ is identified with the center of $\text{SL}(2,\mathbbm{Z})$, that is, precisely the part of $\text{SL}(2,\mathbbm{Z})$ which leads to a linear action on $\tau$, $z$ and $\tilde{z}$. This was to be expected, given that the $\tilde{G}$ action on $T$ is linear. The action of $\tilde{\mathcal Q}$, on the other hand, involves a ``lattice shift'': the prefactor that arises from the quasi-periodicity of the theta function with index 1/2 is canceled by the non-trivial transformation of the exponential pre-factors $e^{2\pi i d\cdot T}$ and $e^{2\pi i d\cdot T+y}$ in a way that leaves either theta function term in Eq.~\eqref{eq:theta2} invariant separately.

Finally, let us speculate on the reason for why the theta functions appear. The only non-trivial $\text{SL}(2,\mathbbm{Z})$ transformation which originates from the group $\tilde{G}$ is $-\mathbbm{1}_2$. Once we accept the pre-factors $e^{2\pi i d\cdot T}$ and $e^{2\pi i d\cdot T+y}$, the appearance of the Jacobi $\vartheta$-function in the $\tilde{G}$ invariant prepotential is not too surprising. We know that the prepotential has to be holomorphic, and invariance under $\tilde{\mathcal{Q}}$ made use of quasi-periodicity (with index $m=1/2$) of this holomorphic function. Using the theta decomposition~(see, for example \cite{Cheng:2017usy}) for a holomorphic function $\xi(z,\tau)$,
\begin{align}
\xi(z,\tau)=\sum_{r~\text{mod}~2m}h_r(\tau)\theta_{m,r}(z,\tau)\qquad\text{with}\qquad \theta_{m,r}(z,\tau)=\sum_{k\in\mathbbm{Z}}e^{2\pi i [(2km+r)z+\frac{1}{4m}(2km+r)^2\tau]}\,,
\end{align}
and setting $m=1/2$, we find that $\xi(z,\tau)$ is $\vartheta(z,\tau)$ times a function $h(\tau)$ of the modular parameter. Since no group element in $\tilde{G}$ induces a non-trivial transformation in $\tau$, we cannot fix $h(\tau)$ from such a symmetry argument, but we know from our calculation that it comes out to be trivial. From a completely different point of view, it was argued in Ref.~\cite{Huang:2015sta} that the topological string partition function of elliptically fibered Calabi-Yau manifolds can be expressed in terms of (a quotient of) even weak Jacobi forms. As we argue in Appendix~\ref{app:EllFib}, for the class of models we study, that is, complete intersections in products of projective ambient spaces which have a GLSM charge matrix that ensures the existence of infinitely many flops as worked out in~\cite{Brodie:2021toe}, the cases with $m_1m_2\leq4$ always lead to Calabi-Yau manifolds that have an elliptic fibration (with a section or a multi-section). In the appendix, we also explain how to find the Kollar divisor in each case. Therefore, we can find a basis change of the K\"ahler cone generators that makes the base and fiber classes (which are a divisor and a curve class given as a complete intersection, respectively) more apparent. By carrying out this base change, it should be possible to match our expressions to the ones given in Ref.~\cite{Huang:2015sta}. The details of this match are, however, beyond the scope of this paper.

We give an overview of how many CICYs (up to $h^{1,1}=7$) realize the various cases discussed above in Table~\ref{tab:CICYcases}.
\begin{table}[t]
\centering
\begin{tabular}{|c|c|c|c|c|c|}\hline
$h^{1,1}(X)$&\# CICYs&$\geq$ 2 involutions&$m_1m_2<4$&$m_1m_2=4$&$m_1m_2>4$\\\hline\hline
3&155&81&38&23&20\\\hline
4&425&186&90&96&0\\\hline
5&837&179&124&55&0\\\hline
6&1140&74&74&0&0\\\hline
7&1112&38&38&0&0\\\hline
\end{tabular}
\caption{The number of CICYs, CICYs with at least two involutions and numbers which realize the three cases for $m_1m_2$ for Picard numbers $h^{1,1}(X)\leq 7$.}
\label{tab:CICYcases}
\end{table}
The table shows larger values of $m_1m_2$ become rarer as $h^{1,1}(X)$ increases, to the extent that $h^{1,1}(X)=7$ CICYs with at least two flop boundaries all satisfy $m_1m_2<4$ and, hence, lead to finite symmetry groups only. We do not know if this ``decline" of the group order with increasing $h^{1,1}(X)$ is a feature of the CICY construction or a feature of CY manifolds more generally. In this context, it would be interesting to analyze other classes of CY manifolds, such as those obtained from the Kreuzer-Skarke classification.

\subsection{Examples}
\subsubsection{Example with three K\"ahler moduli and two isomorphic flops}
\label{sec:ExampleCICY7880}
We would like to discuss some example manifolds with two symmetry generators and modularity. The smallest  Picard rank which allows for such examples is three and a relevant example at $h^{1,1}(X)=3$ is the degree $(2,2,3)$ hypersurface in $\mathbb{P}^1\times\mathbb{P}^1\times\mathbb{P}^2$, which is the CICY with number 7880. Its K\"ahler cone is the positive octant, $\mathcal K=\{(t^1,t^2,t^3)\,|\, t^i\geq 0\}$ and its intersection form
\begin{align}
 \kappa=\lambda_{ijk}t^it^jt^k=18t_1t_2t_3+6t_1{t_3}^2+6t_2{t_3}^2
\end{align}
indicates the presence of two isomorphic flop boundaries at $\{t^1=0\}$ and $\{t^2=0\}$ (while $\{t^3=0\}$ is the end of the effective cone) with associated involutions generated by
\begin{align}
 M_1=\left(
\begin{array}{ccc}
 -1 & 2 & 3 \\
 0 & 1 & 0 \\
 0 & 0 & 1 \\
\end{array}
\right)\,,\qquad
M_2=\left(
\begin{array}{ccc}
 1 & 0 & 0 \\
 2 & -1 & 3 \\
 0 & 0 & 1 \\
\end{array}
\right)\,.
\end{align}
These shows that $(m_1,m_2)=(2,2)$, so this is indeed a case which leads to modularity. Comparison with the general form~\eqref{eq:M1M2gen} of the matrices further shows that $U_1=(3,0)^T$, $U_2=(0,3)^T$, which means $\delta_1=\delta_2=3d_3$. Hence, the $G$-invariant functions $\psi_d^G$ are of the form~\eqref{eq:theta2} with
\begin{align}
\begin{array}{rl}
 \tau&=12d_3(\km_1+\km_2)\\
 z&=2(d_1-d_2)(\km_1+\km_2)+3d_3(\km_1-\km_2)\\
 \tilde{z}&=z-12d_3 \km_1\\
 y&=(-2d_1+2d_2+3d_3)\km_1\,.
 \end{array}
\end{align}

We present the Coxeter diagram for this example in Figure~\ref{fig:Coxeter} on the left.

\subsubsection{Example with five K\"ahler moduli and five isomorphic flops}
\label{sec:HulekVerrill}
As another example we discuss the complete intersection CY defined by two equations of multi-degree $(1,1,1,1,1)$ inside $(\mathbbm{P}^1)^5$, which is the CICY with number 7447. This is the mirror dual~\cite{Candelas:2021lkc} of the Hulek-Verrill manifold~\cite{Hulek:2005aaa}, which has recently received attention in the context of the study of rank two attractor points~\cite{Candelas:2019llw} and its relation to Feynman loop integrals~\cite{Bonisch:2020qmm}. Using the techniques from Appendix~\ref{app:EllFib}, we see that it has isomorphic flops along all five K\"ahler cone boundaries, and all corresponding $M_i$ have $m_i=1$. Hence, the subgroup $G$ of the full symmetry group that is generated by any two of the reflections $G=\langle\mathcal{M}_i,\,\mathcal{M}_j\rangle$ with $i\neq j$ is finite and the prepotential is not modular. In fact, from Table~\ref{tab:finitegroups} we see that $Q^3=(M_i M_j)^3=\mathbbm{1}$. In particular, we can easily write down all the words of $G$ and the sum over the group elements in~\eqref{psidef} explicitly. 

However, if we consider the subgroup generated by three of the $\mathcal{M}_i$, say $G=\langle \mathcal{M}_1,\,\mathcal{M}_2,\,\mathcal{M}_3\rangle$, we can write down infinitely many words, e.g.\ of the type $(\mathcal{M}_1 \mathcal{M}_2 \mathcal{M}_3)^k$. These contain entries that are quadratic, linear, and constant in $k$ and could hence also lead to modular behavior. A detailed study of this is beyond the scope of the current paper, which focuses on the subgroups generated by two involutions, but it would be very interesting to understand the implications of the involutions and the resulting extended K\"ahler cone structure for the mirror. The Coxeter diagram corresponding to this case is given in Figure~\ref{fig:Coxeter} on the right. 

%%%%%%%%%%%%%%%%%%%%%%%%%%%%%%%%%%%%%%%%%%%%%%%%%%%%%%%%%%%%%%%%%%%%%%%%%%%%
\section{Coxeter groups, triangle groups, and reflection symmetries}
\label{sec:CoxGroups}
A Coxeter group is defined as a group with presentation
\begin{align}
W=\langle \mathcal{M}_1,\;\mathcal{M}_2,\;\ldots,\mathcal{M}_r~~|~~(\mathcal{M}_i \mathcal{M}_j)^{c_{ij}}=\mathbbm{1} \rangle\,.
\end{align}
Here, $\mathcal{M}_i$ are generators, $c_{ij}=c_{ji}$, $c_{ii}=1$, and $c_{ij}\geq2$ for $i\neq j$. The condition $c_{ii}=1$ implies that $(\mathcal{M}_i\mathcal{M}_i)^1=\mathbbm{1}$, so all generators $\mathcal{M}_i$ are involutions. If there is no relation of the form $(\mathcal{M}_i\mathcal{M}_j)^{c_{ij}}$, one sets $c_{ij}=\infty$. The symmetric matrix $C$ with entries $c_{ij}$ is called the Coxeter matrix. Conversely, any symmetric matrix with ones on the diagonal and entries $2,3,\ldots,\infty$ otherwise defines a Coxeter group. The pair $(W, \Gamma)$ where $W$ is the Coxeter group and $\Gamma=\{\mathcal{M}_1,\mathcal{M}_2,\ldots,\mathcal{M}_r\}$ are the generators is called a Coxeter system. 

A Coxeter group can be represented by its Coxeter diagram which is constructed from the Coxeter matrix as  follows. First, each generator is represented by a node. Each two nodes $i$ and $j$ are connected with an edge labeled by $c_{ij}$ if $c_{ij}\geq3$; in the case where $c_{ij}=3$ the edge label is omitted. The Coxeter matrix $C_{ij}$ is closely related to the Schl\"afli matrix $S$ with entries $s_{ij}=-2\cos(\pi/c_{ij})$. If the eigenvalues of $S$ are all positive, the Coxeter group is finite. If all eigenvalues are non-negative (and at least one zero), the Coxeter group is affine. Otherwise, the Coxeter group is of indefinite type. Indefinite Coxeter groups with $1$ negative and $r-1$ positive eigenvalue are sometimes called hyperbolic Coxeter groups.

Coxeter~\cite{Coxeter} studied these groups originally in the context where the involutions $M_i$ are reflections along hyperplanes. A canonical representation of a Coxeter group is then in terms of reflection matrices. This representation is constructed by associating to $\mathcal{M}_i$ a basis $\mathcal{M}_i\mapsto e_i$ of $\mathbbm{R}^n$ and by defining a bi-linear form $\mathcal{I}$ and reflections $\sigma_i$ in $\mathbbm{R}^n$ via the Schl\"afli matrix
\begin{align}
S_{ij}:=\mathcal{I}(e_i,e_j)=-2\cos\frac{\pi}{c_{ij}}=s_{ij}\,,\qquad \sigma_i(v)=v-s_{ij} e_i\,.
\end{align}
If $c_{ij}=\infty$, we set $s_{ij}=-2$. Then, the representation $W\to\text{GL}(n,\mathbbm{R})$ defined by $\mathcal{M}_i\mapsto\sigma_i$ is faithful. Moreover, it can be checked that the reflections $\sigma_i$ are isometries of the bi-linear form $\mathcal{I}$, so $\mathcal{I}(\sigma_i(v),\sigma_j(w))=\mathcal{S}(v,w)$ for all $v,w\in\mathbbm{R}^n$.

The question of when a group generated by involutions that are not necessarily reflections along hyperplanes) is a Coxeter group is more difficult, but criteria (like the exchange condition) that describe the behavior of reduced words, are known. Vinberg generalized the concept of reflecting along hyperplanes to involutions across polyhedral cones~\cite{Vinberg:1971aaa}. This is precisely the setup we need for our discussion. 

\begin{figure}[t]
\centering
\includegraphics[width=.55\textwidth]{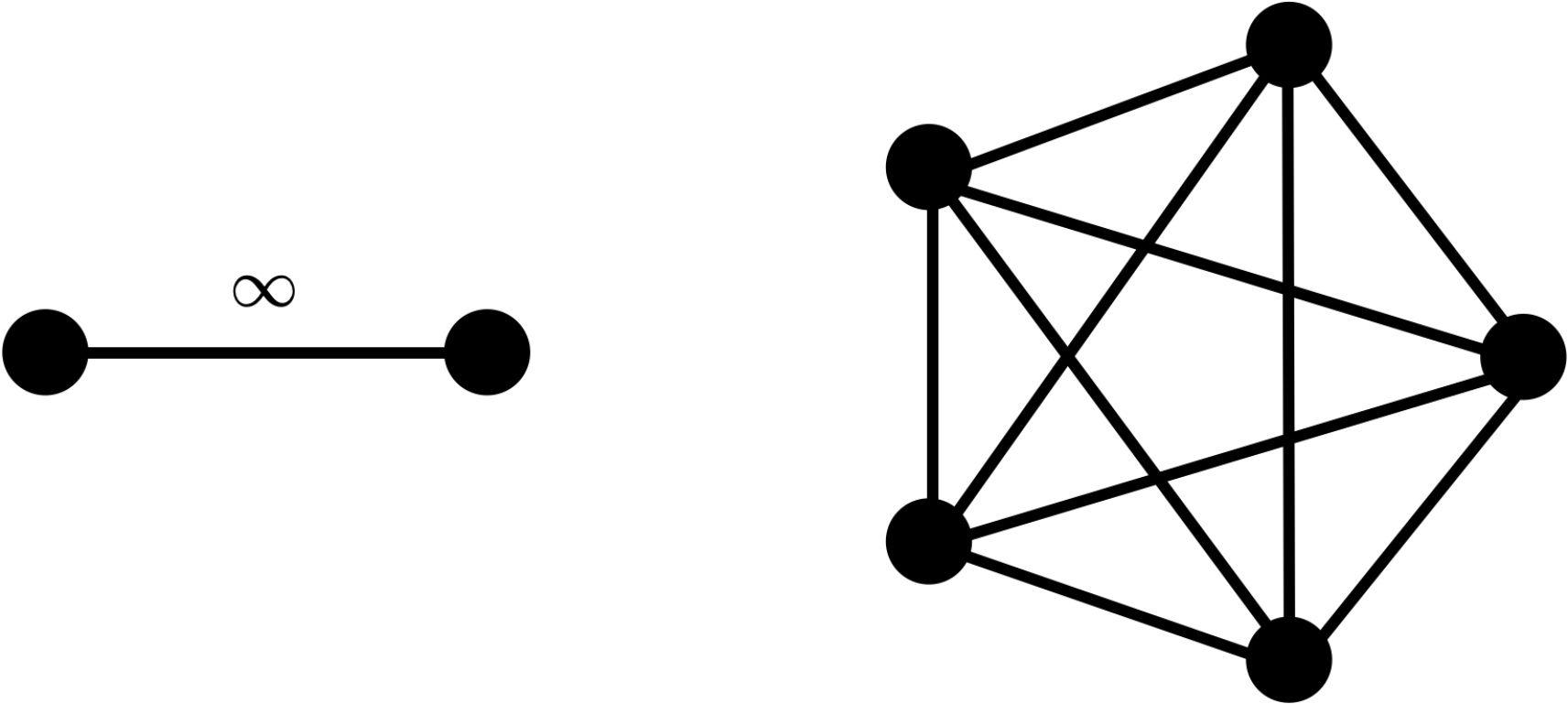}
\caption{Coxeter graphs of the groups $G$ for the examples discussed in this paper. Left: Coxeter diagram of CICY 7863 from the example in Section~\ref{sec:ExamplePicard2} and CICY 7880 from the example in Section~\ref{sec:ExampleCICY7880}. Right: Coxeter diagram of CICY 7447, the mirror of the Hulek-Verrill manifold, from the example in Section~\ref{sec:HulekVerrill}.}
\label{fig:Coxeter}
\end{figure}

Vinberg starts from involutions $\mathcal{M}_i$ of the form~\eqref{eq:ReflectionVector} and shows that the Cartan matrix \footnote{The matrices constructed this way are the Cartan matrices of semisimple Lie algebras, where the reflections are those along the simple roots that generate the Weyl group.} $A$, whose columns are given by the vectors $u_i$ in~\eqref{eq:ReflectionVector}, satisfies $u_{ij}=0$ if $u_{ji}=0$. This has been used in Section~\ref{sec:hGenericGroupStructure} to argue that the case $m_1m_2=0$ implies $m_1=m_2=0$. Vinberg also shows that either $u_{ij}\geq4$ (in which case the entry $c_{ij}$ in the Coxeter matrix is $c_{ij}=\infty$) or $u_{ij}u_{ji}=s_{ij}s_{ji}$. This precisely reproduces Table~\ref{tab:finitegroups}. From this discussion, we see that the groups $G$ (or $\tilde{G}$) acting on the Mori cone (or K\"ahler cone) are Coxeter groups with generators $\mathcal{M}_i$. Their rank $r$ corresponds to the number of facets of the K\"ahler cone across which isomorphic flops occur.

The boundary case $m_im_j=4$ has a geometric realization as reflections along hyperplanes as studied by Coxeter, while $m_im_j>4$ corresponds to more general reflections as considered by Vinberg. In either case, we have $c_{ij}=\infty$ for $i\neq j$ and the Schl\"afli matrix has only entries of 2 (positive on the diagonal, negative elsewhere). The eigenvalues of this matrix are $\lambda_1=\lambda_2=\ldots=\lambda_{r-1}=4$ and $\lambda_r=r-2$. Hence, the case $r=2$, corresponds to an affine Coxeter algebra called $\tilde{I}_1$ (see left part of Figure~\ref{fig:Coxeter}), while the cases with $r>2$ are hyperbolic Coxeter groups. Thus, in these cases, the Coxeter groups are given by fully connected graphs where each edge carries the label $\infty$. These are precisely the universal Coxeter groups, given by the free product of $r$ $\bZ_2$ factors. For the examples with two and five generators studied earlier, the Coxeter graphs are shown in Figure~\ref{fig:Coxeter}. In this paper, we have focused on (sub-)groups $\tilde{G}$ with two generators, so all the infinite reflection groups in this paper are isomorphic to the affine case $\tilde{I}_1$. For such cases, terminology for CY moduli spaces can be translated into terminology for Coxeter groups. Specifically, under the group isomorphism $\tilde{G}\rightarrow \tilde{I}_1$, the K\"ahler cone is identified with the fundamental chamber, the K\"ahler cone walls that admit flops to isomorphic CYs are identified with the walls of the fundamental chamber, and the extended K\"ahler cone is identified with the Tits cone.

Next, we want to briefly study the group elements of a Coxeter group. In general, two or more words can represent the same group element (irrespective of their lengths). This is even true for reduced words, which means that consecutive inverses are deleted. In the context of Coxeter groups, each generator is self-inverse, which means that in reduced words, the same letter does not appear more than once consecutively. For universal Coxeter groups, each reduced word corresponds to a unique group element. Hence, each word $w$ corresponding to a $g\in W$ can be written as $\mathcal{M}_{i_1}\mathcal{M}_{i_2}\cdots \mathcal{M}_{i_n}$ with $i_k\neq i_{k+1}$. This is reflected (no pun intended) in our choice of generators $\mathcal{S}=\mathcal{M}_1$ and $\mathcal{Q}=\mathcal{M}_1\mathcal{M}_2$ for the group $G$.

Finally, let us briefly comment on the relation to triangle groups. Triangle groups are groups which are realized by reflections along the sides of a triangle. They are specified in terms of three integers $(l,m,n)$ which correspond to angles $\pi/l$, $\pi/m$, $\pi/n$ of the triangle. If $1/l+1/m+1/n=1$, the triangles tessellate the Euclidean plane, if $1/l+1/m+1/n>1$ they tessellate the unit sphere, and if $1/l+1/m+1/n<1$ they tessellate the hyperbolic plane. As mentioned above, the non-finite cases with $m_am_b\geq 4$ and $r\geq3$ correspond to hyperbolic Coxeter groups. These are related to the hyperbolic triangle groups (for $r=3$) and their generalizations: The integers $(k,l,m)$ correspond to the entries $c_{1,2}$, $c_{1,3}$ and $c_{2,3}$ of the Coxeter matrix. This means that the general case, which leads to universal Coxeter groups, corresponds to the triangle group $(\infty,\infty,\infty)$.

\begin{figure}
\centering
\includegraphics[width=.35\textwidth]{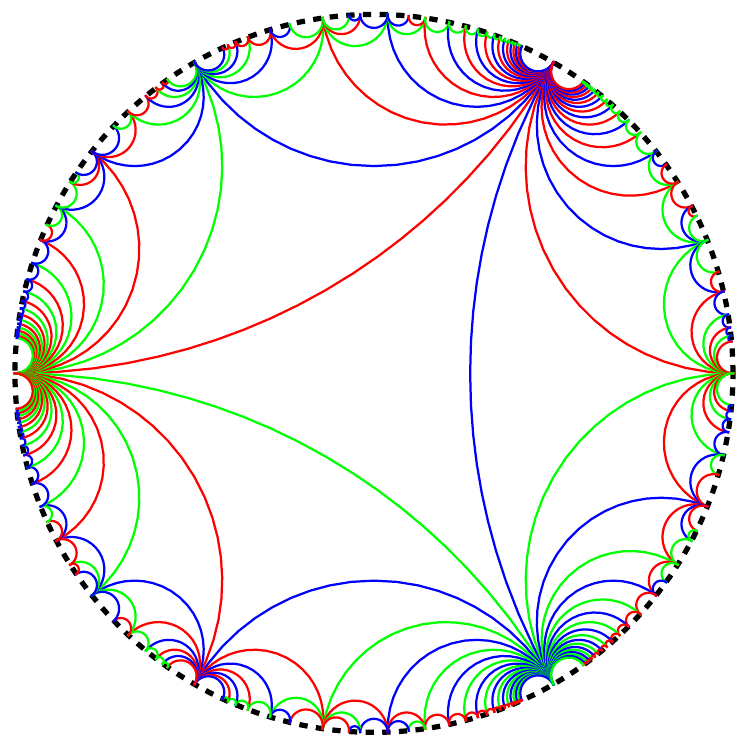}
\caption{The triangle group $(\infty,\infty,\infty)$ on the Poincar\'e disk, which corresponds to the universal Coxeter group of rank 3. In the context of the infinite flop examples for Picard rank 3 CYs discussed here, each arc corresponds to a flop wall. Arcs with the same color are reflection images of one another.}
\label{fig:PoincareDisk}
\end{figure}

As can be seen from the examples discussed in Section~\ref{sec:HigherPicardRank}, or just by noting how $(\mathcal{Q}^T)^k$ behaves, the $\tilde{G}$-images of the original CICY K\"ahler cone become thinner and thinner slivers (in the cohomology basis of the CICY K\"ahler cone, which is the positive octant), accumulating towards the boundaries of the extended K\"ahler cone. They can be resolved better in hyperbolic space, for example by plotting them on the Poincar\'e disk, whose boundary represents infinity and where straight lines are arcs (or diameters of the disk). We plot the triangle group $(\infty,\infty,\infty)$ in Figure~\ref{fig:PoincareDisk}. Under the isomorphism between Coxeter groups and reflections along flop walls of the K\"ahler cone, each arc corresponds to one K\"ahler cone wall. Walls that are identified under the group $\tilde{G}$ are plotted in the same color.

%%%%%%%%%%%%%%%%%%%%%%%%%%%%%%%%%%%%%%%%%%%%%%%%%%%%%%%%%%%%%%%%%%%%%%%%%%%%
\section{Conclusions}
\label{sec:Conclusions}
In this paper, we have studied CY three-folds with isomorphic flops, the resulting reflection symmetries and their implications for the Gopakumar-Vafa (GV) invariants and the K\"ahler moduli prepotential. We have seen that such isomorphic flops lead to symmetry groups $\tilde{G}$ on the K\"ahler moduli space, and dual symmetry groups $G$ acting on curves, which are isomorphic to Coxeter groups. These groups can be of finite or infinite order, depending on the manifold, and the rank of the Coxeter group is given by the number of K\"ahler cone facets across which flops occur. The key observation is that GV invariants $n_d$ for non-flopping curve classes $d$ are $G$-invariant, that is, $n_{gd}=n_d$ for all $g\in G$. This means the (non-flopping part of the) instanton prepotential can be written in terms of suitable $\tilde{G}$ invariant functions, as defined in Eq.~\eqref{psidef}. To our knowledge, these functions, which are invariant under certain representations of Coxeter groups, have not been studied before.

We have analyzed in some detail the case of $h^{1,1}(X)=2$ and isomorphic flops across both facets of the K\"ahler cone which leads to a symmetry group $G$ controlled by two integers $m_1$ and $m_2$. These integers can be computed from the triple intersection numbers via Eq.~\eqref{eq:m1m2def}. For Picard rank two these numbers always satisfy $m_1m_2>4$ and this leads to a symmetry group $\tilde{G}\cong\mathbb{Z}_2\rtimes\mathbb{Z}$, isomorphic to a universal Coxeter group of rank two. In this case, the associated $\tilde{G}$ invariant functions are non-modular.

It turns out, for Picard rank $h^{1,1}(X)>2$ and two isomorphic flop boundaries, all three cases, $m_1m_2<4$, $m_1m_2=4$ and $m_1m_2>4$ can be realized by appropriate CY manifolds. Cases with $m_1m_2<4$ lead to a finite Coxeter group $\tilde{G}$, while for $m_1m_2\geq 4$ the group $\tilde{G}$ is a universal Coxeter group of rank two. If $m_1m_2>4$ the $\tilde{G}$ invariant functions are non-modular, as they were for Picard rank two. The interesting new case is the limiting one, $m_1m_2=4$, where the $\tilde{G}$ invariant functions can be expressed in terms of Jacobi theta functions. We also argued that the appearance of theta functions is related to the presence of elliptic fibrations in the underlying CYs of the type studied in Ref.~\cite{Brodie:2021toe}.

For manifolds with more than two isomorphic flop boundaries the structure is more complicated. While the group $G$ is isomorphic to a Coxeter group, it is not clear how to write down the group elements systematically. It is, therefore, difficult to work out the form of the $\tilde{G}$ invariant functions more explicitly. However, we have argued, for the example of the Hulek-Verrill manifold, that modularity can arise for (sub-)groups $\tilde{G}$ with more than two generators. Moreover, the example of the Hulek-Verrill manifold illustrates that the full symmetry group can have infinitely many elements even though the rank two subgroup has $m_1=m_2=1$ and is thus finite. The detailed analysis of manifolds with such larger symmetry groups and their prepotentials is an interesting direction for future work.

It has been argued in Ref.~\cite{Brodie:2021ain} that the symmetry $G$ needs to be gauged in order to avoid a conflict with the swampland distance conjecture. (The infinite-length geodesic connecting the points $\tilde{g}t$, where $\tilde{g}\in\tilde{G}$, is not associated with a tower of massless particles.) Hence, the low-energy theory is based on the moduli space $\mathcal{K}_{\rm ext}/\tilde{G}$ and $\tilde{G}$ does not actually appear as a symmetry of this theory. Nevertheless, we can think about the ``upstairs" effective theory defined on $\mathcal{K}_{\rm ext}$ which we expect to be $\tilde{G}$ invariant. A $\tilde{G}$ invariant scalar potential in this theory has stationary points at the fixed loci of the $\tilde{G}$ action and we know from our discussion that these fixed loci are precisely the boundaries of the K\"ahler cone, that is, the flop loci. This observation might well have implications for moduli stabilisation, although more precise statements require further study, for example in the context of type II models with D-branes and flux. This is beyond the scope of the present paper but it may be an interesting avenue for further study.

\subsection*{Acknowledgments}
We thank Callum Brodie, Sarah Harrison, Albrecht Klemm, Wolfgang Lerche, Guglielmo Lockhart, and Thorsten Schimmanek for useful discussions. The work of FR is supported by the NSF grant PHY-2210333, by the NSF under Cooperative Agreement PHY-2019786 (The NSF AI Institute for Artificial Intelligence and Fundamental Interactions), and by startup funding from Northeastern University.

\appendix
%%%%%%%%%%%%%%%%%%%%%%%%%%%%%%%%%%%%%%%%%%%%%%%%%%%%%%%%%%%%%%%%%%%%%%%%%%%%
\section{Appendix: Special symmetry groups and elliptic fibrations}
\label{app:EllFib}
In this appendix we would like to study the CICYs with $m_1m_2\leq4$ somewhat more systematically and we will argue that these cases lead to elliptic fibrations. To do so, we distinguish the two types of CICYs that lead to infinitely many isomorphic flops, as identified in Ref.~\cite{Brodie:2021toe}, separately. As we will show, type 1 CICYs lead to $(m_1,m_2)\in\{(1,1),(2,2)\}$ only, while type 2 CICYs allow for $(m_1,m_2)\in \{(1,4),(4,1),(2,2)\}$. The cases $(m_1,m_2)\in\{(1,2),(1,3)\}$, which also appear in Table~\ref{tab:finitegroups}, are not realized by CICYs. One example of this is the mirror dual of the Hulek-Verrill manifold as discussed in Section~\eqref{sec:HulekVerrill}. It is intriguing that theta functions also appear for this manifold, albeit only once three or more reflections are considered.

\subsection{Type 1}
For type 1 CICYs, the configuration matrix is of the form
\begin{align}
\left[
\begin{array}{c|cccccc}
\mathbbm{P}^n & 1&1&\ldots&1&0&\ldots\\
\vec{\mathbbm{P}} & \vec{q}&\vec{q}&\ldots&\vec{q}&\vec{q}_{n+1}&\ldots
\end{array}
\right]
\end{align}
The reflection matrix $\mathcal{M}_1$ is given by~\eqref{eq:tildeMReflect} with
\begin{align}
\label{eq:M1Type1}
u=(2,-n\vec{q}^{\;T})\,.
\end{align}
Now we want this row to contain an entry $m_1\leq4$, such that, when combined with the second flop that will generate the infinite flop chain, we can have $m_1 m_2 \leq4$. Since $n$ is the dimension of the ambient space $\mathbbm{P}^n$ factor, we know that $n\geq1$. Since all $\vec{q}$ have to be the same in type 1 flops, they actually cannot be combined with a type 2 flop in a sensible way: In a type 2 flop, the entries that occur more than once are 0's and 1's. If all $\vec{q}$ contain a zero row, this means that the configuration matrix is block-diagonal and hence the complete intersection is a direct product. More precisely, we would get two linear equations in $\mathbbm{P}^1$ times the CICY defined by $[\vec{\mathbbm{P}}~|~\vec{q}_{n+1}~~\ldots]$, which has no solution for generic CS, or just reproduces the original CICY over special points in CS space. Either one is not interesting. If the $\vec{q}$ contain a row with 1's, we already get an entry $m_2$ that is at least $m_2=2+2\cdot2=6$ according to~\eqref{eq:M1Type2}. So we can focus on the case where the second flop is of the same type as the first flop.

In such a setup, because of~\eqref{eq:M1Type1}, we can focus on $1\leq n\leq 3$, and possible configuration matrices are of the form
\begin{align}
\label{eq:InfFlop1}
\left[
\begin{array}{c|cccc}
\mathbbm{P}^1 & 1&1&0&\ldots\\
\mathbbm{P}^1 & 1&1&0&\ldots\\
\vec{\mathbbm{P}} & \vec{q}&\vec{q}&\vec{q}_{3}&\ldots
\end{array}
\right]\,,
\quad
\left[
\begin{array}{c|ccccc}
\mathbbm{P}^2 & 1&1&1&0&\ldots\\
\mathbbm{P}^2 & 1&1&1&0&\ldots\\
\vec{\mathbbm{P}} & \vec{q}&\vec{q}&\vec{q}&\vec{q}_{4}&\ldots
\end{array}
\right]\,,\hspace{2cm}
\end{align}
giving rise to $(m_1,m_2)=(1,1)$ and $(m_1,m_2)=(2,2)$, respectively.

We can argue next that CICYs of the type~\eqref{eq:InfFlop1} are always elliptically fibered. To do so, we use a criterion due to Oguiso and Kollar. This comes down to showing that there exists an effective divisor $D_*$ such that $D_*.C\geq0$, $D_*.D_*\neq0$, and $D_*.D_*.D_*=0$ for any algebraic curve $C\subset X$. So in essence we get a basis of divisors by pullbacks from the ambient space, compute their triple intersection numbers, and look for a positive linear combination $D_*=\sum_{i=1}^{h^{11}} a_i D_i$ such that $D_*^2\neq0$ and $D_*^3=0$ ($D.C\geq0$ is automatic for effective divisors and curves). For CICYs of the first form, the divisors $D_1$ and $D_2$ corresponding to the two ambient space $\mathbbm{P}^1$ factors are good candidates. They automatically have $D^3=0$. However, they also have $D^2=0$. This can be easily cured by taking the linear combination $D_*=D_1+D_2$. Now $D_*^2=(D_1+D_2)^2=2D_1.D_2$ fixes a point in $\mathbbm{P}^1\times\mathbbm{P}^1$ on the CY, and will intersect some other effective divisor with non-zero intersection number, i.e., we have an elliptic fibration. For CICY's of the last type, either of the divisors $D_1$ and $D_2$ satisfies $D_i^2\neq0$ and $D_{i}^3=0$ on dimensional grounds. The other cases cannot be treated in generality.

\subsection{Type 2}
For type 2 CICYs, the configuration matrix is of the form
\begin{align}
\left[
\begin{array}{c|cccccc}
\mathbbm{P}^n &2&1&\ldots&1&0&\ldots\\
\vec{\mathbbm{P}} & \vec{q}_1&\vec{q}_2&\ldots&\vec{q}_n&\vec{q}_{n+1}&\ldots
\end{array}
\right]
\end{align}
The reflection matrix $\mathcal{M}_1$ along the $\mathbbm{P}^n$ direction will be given by~\eqref{eq:tildeMReflect} with
\begin{align}
\label{eq:M1Type2}
u=(2,\;-\vec{q}_1^{\,T}-2\sum_{i=2}^n \vec{q}_i^{\,T})\,.
\end{align}
In fact, since we want infinitely many flops, we need a second direction along which we can flop (say the second ambient space factor. That means the configuration matrix is of the form
\begin{align}
\left[
\begin{array}{c|ccccccc}
\mathbbm{P}^{n} &2&1&\ldots&1&0&\ldots\\
\mathbbm{P}^{\tilde{n}} &q_1&q_2&\ldots&q_n&q_{n+1}&\ldots\\
\vec{\mathbbm{P}} & \vec{q}_1^{\;'}&\vec{q}_2^{\;'}&\ldots&\vec{q}_n^{\;'}&\vec{q}_{n+1}^{\;'}&\ldots
\end{array}
\right]
\end{align}
The part with $\vec{\mathbbm{P}}$ is irrelevant for our subsequent discussion, so we will focus on the first two rows. We want to find entries $q_1,q_2,\ldots$ such that the resulting rows have entries $m_1$ and $m_2$ with $m_1m_2\leq4$. In principle we can choose $q_1,q_2,\ldots$ such that we get a type 1 or a type 2 flop. For a type 1 flop, we need the $q_i$ to be either 1 or 0. If they are 1, then the rest of the matrix has to be the same for all positions where a 1 occurs. This means in particular that $q_1=0$, since there is only a single entry $2$ in the first row. Similarly, since the $m_1$ and the $m_2$ should not ``miss'' each other, we cannot use the entries $q_{n+1}$, $q_{n+2}$, etc.\ and can set these to zero as well. In other words, we can choose a subset of $q_2$ to $q_n$ to be 1, and the rest is zero. However, as we can see from~\eqref{eq:M1Type2}, the entries $q_2$, $q_3$, $\ldots$, $q_n$ come with a factor of 2, so $m_1m_2\leq4$ would require only one of these being non-zero, which would mean an ambient space $\mathbbm{P}^0$. So we cannot generate an infinite flop chain with $m_1 m_2\leq4$ by using types 1 and types 2. This leaves us with choosing the $q_i$ of the second row to be of type 2, i.e., a single two, and a bunch of ones and zeros. By choosing the $q_i$ appropriately, we can engineer examples with $m_1m_2=4$ in three different ways:

The first way is to set $q_1=2$, $q_i=0$ for $i=2,\ldots,n$, and choosing the remaining $q_i$ for $i\geq n+1$ to be either 1 or 0 (to obtain a type 2 flop). This gives $(m_1,m_2)=(2,2)$ and a configuration matrix of the form
\begin{align}
\left[
\begin{array}{c|ccccccc}
\mathbbm{P}^n &2&1&\ldots&1&0&\ldots\\
\mathbbm{P}^{\tilde{n}} &2&0&\ldots&0&q_{n+1}&\ldots\\
\vec{\mathbbm{P}} & \vec{q}_1^{\;'}&\vec{q}_2^{\;'}&\ldots&\vec{q}_n^{\;'}&\vec{q}_{n+1}^{\;'}&\ldots
\end{array}
\right]
\end{align}

\bigskip

The second way is to set $q_1=1$, $q_i=0$ for $i=2,\ldots,n$, and choosing one of the remaining $q_i$ for $i\geq n+1$ to be 2 and all others to be either 1 or 0 (to obtain a type 2 flop). This gives $(m_1,m_2)=(1,4)$ and a configuration matrix of the form
\begin{align}
\left[
\begin{array}{c|cccccccc}
\mathbbm{P}^n &2&1&\ldots&1&0&0&\ldots\\
\mathbbm{P}^{\tilde{n}} &1&0&\ldots&0&2&q_{n+2}&\ldots\\
\vec{\mathbbm{P}} & \vec{q}_1^{\;'}&\vec{q}_2^{\;'}&\ldots&\vec{q}_n^{\;'}&\vec{q}_{n+1}^{\;'}&\vec{q}_{n+2}^{\;'}&\ldots
\end{array}
\right]
\end{align}

\bigskip

The third way is to to set $q_1=0$, precisely one of the $q_i$, $i=2,\ldots,n$ to one, and choosing one of the remaining $q_i$ for $i\geq n+1$ to be 2 and all others to be either 1 or 0 (to obtain a type 2 flop). This also gives $(m_1,m_2)=(2,2)$ and a configuration matrix of the form
\begin{align}
\left[
\begin{array}{c|ccccccccc}
\mathbbm{P}^n &2&1&1&\ldots&1&0&0&\ldots\\
\mathbbm{P}^{\tilde{n}} &0&1&0&\ldots&0&2&q_{n+2}&\ldots\\
\vec{\mathbbm{P}} & \vec{q}_1^{\;'}&\vec{q}_2^{\;'}&\vec{q}_3^{\;'}&\ldots&\vec{q}_n^{\;'}&\vec{q}_{n+1}^{\;'}&\vec{q}_{n+2}^{\;'}&\ldots
\end{array}
\right]
\end{align}

\bigskip

As examples for CICYs that realize these three possibilities, we can take the tetra-quadric, CICY 5299 (these even have flops along all ambient space factors), and CICY 6971, respectively.
\begin{align}
X_\text{TQ}\sim\left[\begin{array}{c|c}
\mathbbm{P}^1&2\\\mathbbm{P}^1&2\\\mathbbm{P}^1&2\\\mathbbm{P}^1&2
\end{array}
\right]\,,\qquad
X_{5299}\sim\left[\begin{array}{c|ccc}
\mathbbm{P}^2&2&1&0\\\mathbbm{P}^2&1&0&2\\\mathbbm{P}^2&0&2&1
\end{array}
\right]\,,\qquad
X_\text{6971}\sim\left[\begin{array}{c|ccc}
\mathbbm{P}^2&2&1&0\\\mathbbm{P}^2&0&1&2\\\mathbbm{P}^2&1&1&1
\end{array}
\right]\,.
\end{align}

\bigskip

Note that for all three possibilities, we get $(n+\tilde{n}-1)$ equations that involve the ambient spaces $\mathbbm{P}^{n}\times\mathbbm{P}^{\tilde{n}}$. This is because there are $n$ (resp.\ $\tilde{n}$) non-zero entries in the first (resp.\ second) row, and precisely in one column, both the first and second row have a non-zero entry. Hence we get $(n+\tilde{n}-1)$ equations inside $\mathbbm{P}^{n}\times\mathbbm{P}^{\tilde{n}}$, i.e., a CY one-fold. This is what the authors of Ref.~\cite{Gray:2014fla} call an ``obvious elliptic fibration''.

%%%%%%%%%%%%%%%%%%%%%%%%%%%%%%%%%%%%%%%%%%%%%%%%%%%%%%%%%%%%%%%%%%%%%%%%%%%%
\bibliographystyle{bibstyle}
\bibliography{refs}

\providecommand{\href}[2]{#2}\begingroup\begin{thebibliography}{10}

\bibitem{Aspinwall:1993yb}
P.~S. Aspinwall, B.~R. Greene, and D.~R. Morrison ``{Multiple mirror manifolds
  and topology change in string theory},'' {\em Phys. Lett. B} {\bf 303} (1993)
  249--259 \href{http://www.arXiv.org/abs/hep-th/9301043}{[{\tt
  hep-th/9301043}]}.

\bibitem{Aspinwall:1993nu}
P.~S. Aspinwall, B.~R. Greene, and D.~R. Morrison ``{Calabi-Yau moduli space,
  mirror manifolds and space-time topology change in string theory},'' {\em
  Nucl. Phys. B} {\bf 416} (1994) 414--480
  \href{http://www.arXiv.org/abs/hep-th/9309097}{[{\tt hep-th/9309097}]}.

\bibitem{Greene:1996cy}
B.~R. Greene ``{String theory on Calabi-Yau manifolds},'' in {\em {Theoretical
  Advanced Study Institute in Elementary Particle Physics (TASI 96): Fields,
  Strings, and Duality}} pp.~543--726.
\newblock 6, 1996.
\newblock \href{http://www.arXiv.org/abs/hep-th/9702155}{[{\tt
  hep-th/9702155}]}.

\bibitem{Brodie:2021toe}
C.~Brodie, A.~Constantin, A.~Lukas, and F.~Ruehle ``{Flops for Complete
  Intersection Calabi-Yau Threefolds},''
  \href{http://www.arXiv.org/abs/2112.12106}{[{\tt 2112.12106}]}.

\bibitem{Brodie:2021nit}
C.~R. Brodie, A.~Constantin, A.~Lukas, and F.~Ruehle ``{Geodesics in the
  extended K\"ahler cone of Calabi-Yau threefolds},'' {\em JHEP} {\bf 03}
  (2022) 024 \href{http://www.arXiv.org/abs/2108.10323}{[{\tt 2108.10323}]}.

\bibitem{Brodie:2020fiq}
C.~R. Brodie, A.~Constantin, and A.~Lukas ``{Flops, Gromov-Witten invariants
  and symmetries of line bundle cohomology on Calabi-Yau three-folds},'' {\em
  J. Geom. Phys.} {\bf 171} (2022) 104398
  \href{http://www.arXiv.org/abs/2010.06597}{[{\tt 2010.06597}]}.

\bibitem{Coxeter}
H.~S.~M. Coxeter ``Discrete groups generated by reflections,'' {\em Annals of
  Mathematics} {\bf 35} 588--621.

\bibitem{Vinberg:1971aaa}
{\`{E}}.~B. Vinberg ``{DISCRETE} {LINEAR} {GROUPS} {GENERATED} {BY}
  {REFLECTIONS},'' {\em Mathematics of the {USSR}-Izvestiya} {\bf 5} (oct,
  1971) 1083--1119.

\bibitem{Gendler:2022qof}
N.~Gendler, M.~Kim, L.~McAllister, J.~Moritz, and M.~Stillman
  ``{Superpotentials from Singular Divisors},''
  \href{http://www.arXiv.org/abs/2204.06566}{[{\tt 2204.06566}]}.

\bibitem{Gopakumar:1998ii}
R.~Gopakumar and C.~Vafa ``{M theory and topological strings. 1.},''
  \href{http://www.arXiv.org/abs/hep-th/9809187}{[{\tt hep-th/9809187}]}.

\bibitem{Gopakumar:1998jq}
R.~Gopakumar and C.~Vafa ``{M theory and topological strings. 2.},''
  \href{http://www.arXiv.org/abs/hep-th/9812127}{[{\tt hep-th/9812127}]}.

\bibitem{Candelas:1987kf}
P.~Candelas, A.~M. Dale, C.~A. Lutken, and R.~Schimmrigk ``{Complete
  Intersection Calabi-Yau Manifolds},'' {\em Nucl. Phys.} {\bf B298} (1988)
493.
%%CITATION = NUPHA,B298,493;%%.

\bibitem{Candelas:2021lkc}
P.~Candelas, X.~de~la Ossa, P.~Kuusela, and J.~McGovern ``{Mirror Symmetry for
  Five-Parameter Hulek-Verrill Manifolds},''
  \href{http://www.arXiv.org/abs/2111.02440}{[{\tt 2111.02440}]}.

\bibitem{Hulek:2005aaa}
K.~Hulek and H.~Verrill ``On modularity of rigid and nonrigid calabi-yau
  varieties associated to the root lattice a 4,'' {\em Nagoya Mathematical
  Journal} {\bf 179} (2005) 103--146.

\bibitem{Candelas:1990rm}
P.~Candelas, X.~C. De~La~Ossa, P.~S. Green, and L.~Parkes ``{A Pair of
  Calabi-Yau manifolds as an exactly soluble superconformal theory},'' {\em
  Nucl. Phys. B} {\bf 359} (1991) 21--74.

\bibitem{Cheng:2017usy}
M.~C.~N. Cheng, J.~F.~R. Duncan, and J.~A. Harvey ``{Weight One Jacobi Forms
  and Umbral Moonshine},'' {\em J. Phys. A} {\bf 51} (2018) no.~10, 104002
  \href{http://www.arXiv.org/abs/1703.03968}{[{\tt 1703.03968}]}.

\bibitem{Huang:2015sta}
M.-x. Huang, S.~Katz, and A.~Klemm ``{Topological String on elliptic CY 3-folds
  and the ring of Jacobi forms},'' {\em JHEP} {\bf 10} (2015) 125
  \href{http://www.arXiv.org/abs/1501.04891}{[{\tt 1501.04891}]}.

\bibitem{Candelas:2019llw}
P.~Candelas, X.~de~la Ossa, M.~Elmi, and D.~Van~Straten ``{A One Parameter
  Family of Calabi-Yau Manifolds with Attractor Points of Rank Two},'' {\em
  JHEP} {\bf 10} (2020) 202 \href{http://www.arXiv.org/abs/1912.06146}{[{\tt
  1912.06146}]}.

\bibitem{Bonisch:2020qmm}
K.~B\"onisch, F.~Fischbach, A.~Klemm, C.~Nega, and R.~Safari ``{Analytic
  structure of all loop banana integrals},'' {\em JHEP} {\bf 05} (2021) 066
  \href{http://www.arXiv.org/abs/2008.10574}{[{\tt 2008.10574}]}.

\bibitem{Brodie:2021ain}
C.~R. Brodie, A.~Constantin, A.~Lukas, and F.~Ruehle ``{Swampland conjectures
  and infinite flop chains},'' {\em Phys. Rev. D} {\bf 104} (2021) no.~4,
  046008 \href{http://www.arXiv.org/abs/2104.03325}{[{\tt 2104.03325}]}.

\bibitem{Gray:2014fla}
J.~Gray, A.~S. Haupt, and A.~Lukas ``{Topological Invariants and Fibration
  Structure of Complete Intersection Calabi-Yau Four-Folds},'' {\em JHEP} {\bf
  09} (2014) 093 \href{http://www.arXiv.org/abs/1405.2073}{[{\tt 1405.2073}]}.

\end{thebibliography}\endgroup
\end{document}